\documentclass[10pt]{article}
\usepackage{amsmath}
\usepackage{graphicx}
\usepackage{appendix}
\usepackage{amssymb}

\def\beq#1{\begin{equation}\label{#1}}
\def\eeq{\end{equation}}
\def\beqa#1{\begin{eqnarray}\label{#1}}
\def\eeqa{\end{eqnarray}}

\def\Eq#1{(\ref{#1})}

\def\myfrac#1#2{\left(\frac{#1}{#2}\right)}
\def\comment#1{\relax}
 
\newcommand{\ms}{M_\odot}
\newcommand{\msun}{M_\odot}
 
\begin{document}
\noindent
\textbf{\large Settling accretion onto slowly rotating X-ray pulsars}\\
{\large N. I. Shakura, K. A. Postnov, A. Yu. Kochetkova, L. Hjalmarsdotter}\\
\\
\textit{Sternberg Astronomical Institute, Moscow M.V. Lomonosov State University, Universitetskij pr.13, 119992, Moscow, Russia}\\
\\
\\
This work considers a theoretical model for quasi-spherical subsonic accretion onto slowly rotating magnetized 
neutron stars. In this regime the accreting matter settles down subsonically onto the rotating magnetosphere, forming an extended quasi-static shell. 
The shell mediates the angular momentum transfer to/from the rotating neutron star 
magnetosphere by large-scale convective motions, which for observed pulsars lead to an almost so-angular-momentum rotation law with $\omega \sim 1/R^2$ inside the shell. The accretion rate through the shell is determined by the ability of the plasma to enter the magnetosphere due to Rayleigh-Taylor instabilities while taking cooling into account.
The settling regime of accretion is possible for moderate accretion rates 
$\dot M\lesssim \dot M_*\simeq 4\times 10^{16}$~g/s. At higher accretion rates a free-fall gap 
above the neutron star magnetosphere appears due to rapid Compton cooling, and accretion 
becomes highly non-stationary. 
From observations of spin-up/spin-down rates of quasi-spherically wind accreting equilibrium X-ray pulsars with known orbital periods (like e.g. GX 301-2 and Vela X-1), it is possible to determine the main dimensionless parameters of the model, as well as to estimate the magnetic field on the surface of the neutron star.  
For equilibrium pulsars  with independent measurements of the magnetic field, the model also allows us to estimate the velocity of the stellar wind from the companion without the use of complicated spectroscopic measurements. 
For non-equilibrium pulsars, it can be shown that there exists a maximum possible value of the spin-down rate of the accreting neutron star. From observations of the spin-down rate and the X-ray luminosity in such pulsars (e. g. GX 1+4, SXP 1062 and 4U 2206+54) we are able to estimate a lower limit on the neutron star magnetic field, which in all exemplified cases turns out to be close to the standard one and in agreement with cyclotron line measurements.
The model further explains both the spin-up/spin-down 
of the pulsar frequency on large time-scales and the irregular short-term frequency
fluctuations, which may correlate or anti-correlate with the X-ray luminosity fluctuations, seen in 
different systems.

\section{Introduction}
\label{intro}
X-ray pulsars are highly magnetized rotating neutron stars in close binary systems, accreting matter from a companion star. The companion may be a low-mass star overfilling its Roche lobe, in which case an accretion disc is formed. In the case of a high-mass companion, the neutron star may also accrete from the strong stellar wind, and depending on the conditions a disc may be formed or accretion may take place quasi-spherically. The strong magnetic field (of the order of
$10^{12}-10^{13}$~G) of the neutron star disrupts the accretion flow at some distance from the neutron star surface and forces the accreted matter to funnel down on the polar caps of the neutron star creating hot spots that, if misaligned with the rotational axis, make the neutron star pulsate in X-rays. Most accreting pulsars show stochastic variations in their spin frequencies as well as in their luminosities. Many sources also exhibit long-term trends in their spin-behaviour with the period more or less steadily increasing or decreasing, and in some sources spin-reversals have been observed. (For a thorough review, see e.g. \cite{Bildsten_ea97} and references therein.)

The best-studied case of accretion is that of thin disc accretion \cite{ShakuraSunyaev73}. Here the spin-up/spin-down mechanisms are rather well understood.
For disc accretion the spin-up torque is determined by the specific angular momentum at the inner edge of the disc and can be written in the form 
$
K_{su}\approx \dot M\sqrt{GMR_A}\,
$
\cite{PringleRees72}. For a pulsar the inner radius of the accretion disc is determined by the Alfv\'en radius 
$R_A\sim \dot M^{-2/7}$, so $K_{su}\sim \dot M^{6/7}$, i.e. 
for disc accretion the spin-up torque is weakly (almost linearly) dependent on the accretion rate (X-ray luminosity). 
In contrast, the spin-down torque for disc accretion in the first approximation is independent of $\dot M$: 
$K_{sd}\sim -\mu^2/R_c^3$, where 
$R_c=(GM/(\omega^*)^2)^{1/3}$ is the corotation radius, $\omega^*$ is the neutron star angular frequency and $\mu$ is
the neutron star's dipole magnetic moment. In fact, accretion torques in disc accretion are determined by complicated disc-magnetospheric interactions, see, e.g., 
\cite{GhoshLamb79},\cite{Lovelace_ea95} and the 
discussion in \cite{KluzniakRappaport07}, and correspondingly can have a 
more complicated dependence on the mass accretion rate and other parameters.

Measurements of spin-up/spin-down in X-ray pulsars can be used to evaluate a very important parameter of the neutron star -- its magnetic field. The period of the pulsar is usually close to the equilibrium value $P_{eq}$, which is determined by the total zero torque applied to the neutron star, $K=K_{su}+K_{sd}=0$. So assuming the observed 
value $\omega^*=2\pi/P_{eq}$, the magnetic field of the neutron star in disc-accreting X-ray pulsars can be estimated if $\dot M$ is known. 

In the case of quasi-spherical accretion, which may take place in systems where the optical star underfills its Roche lobe and no accretion disc is formed,
the situation is more complicated. Clearly, the amount and sign of the 
angular momentum supplied to 
the neutron star from the captured stellar wind are important for spin-up or spin-down. To within a numerical factor of the order of 1 (which can be positive or negative, see numerical simulations
by \cite{FryxellTaam88},\cite{Ruffert97}, \cite{Ruffert99}, etc.), 
the torque applied to the neutron star 
in this case should be proportional to $\dot M \omega_B R_B^2$, where $\omega_B=2\pi/P_B$ is the binary orbital angular frequency, $R_B=2GM/(V_w^2+v_{orb}^2)^2$ 
is the gravitational capture (Bondi) radius, $V_w$ is the stellar wind velocity at the neutron star orbital distance, and $v_{orb}$ is the neutron star orbital velocity. In real high-mass X-ray binaries the orbital eccentricity is non-zero, the stellar wind 
is variable and can be inhomogeneous, etc., so $K_{su}$ can be a complicated function of time. The spin-down torque is
even more uncertain, since it is impossible to write down a simple equation like $-\mu^2/R_c^3$ any more
($R_c$ has no meaning for quasi-spherical accretion; for slowly rotating pulsars 
it is much larger than the Alfv\'en radius where the angular momentum transfer from the accreting matter to the magnetosphere actually occurs). For example, using the expression $-\mu^2/R_c^3$
for the braking torque results in a very high ($\geq10^{14}$~G) magnetic field 
for long-period X-ray pulsars. We think this is a result
of underestimating the braking torque.

The matter captured from the stellar wind can accrete onto the neutron star 
in different ways. Indeed, if the X-ray flux from the accreting neutron star 
is sufficiently high, the shocked matter rapidly cools down due to Compton processes
and falls freely toward the magnetosphere. The velocity of motion 
rapidly becomes supersonic, so a shock is formed above the magnetosphere. This regime was considered, e.g., by \cite{Burnard_ea83}. Depending on the 
sign of the specific angular momentum of falling 
matter (prograde or retrograde), the neutron star can spin-up or spin-down. However, 
if the X-ray flux at the Bondi radius is below some value, 
the shocked matter remains hot, the radial velocity of the plasma is subsonic, and the source may enter the 
settling accretion regime. A hot quasi-static shell forms around the magnetosphere \cite{DaviesPringle81}. Due to additional energy release
(especially near the base of the shell), the temperature gradient across the shell
becomes superadiabatic, and large-scale convective motions inevitably appear. 
The convection initiates turbulence, and the motion of a fluid element in the shell 
becomes quite complicated. If the magnetosphere allows plasma entry via instabilities
(and subsequent accretion onto the neutron star), the actual accretion rate
through such a shell is controlled by the magnetosphere (for example, a shell 
can exist, but accretion through it can be weak or even absent altogether). Therefore, on top of the convective motions, the matter acquires a low, on average radial, velocity toward the magnetosphere, and thus subsonic settling is possible. This type of accretion can work only for small X-ray luminosities, $L_x<4\times 10^{36}$~erg/s (see below), 
and is totally different from that considered in the numerical simulations cited above.   
If a shell is present, its interaction with the rotating 
magnetosphere can lead to spin-up or spin-down of the neutron star, depending 
on the sign of the difference of the angular velocity between the 
accreting matter and the magnetospheric 
boundary. Thus, in the settling accretion regime, both spin-up or spin-down 
of the neutron star is possible, even if the sign of the specific angular momentum of the
captured matter is always prograde. The shell here mediates the angular momentum transfer
to or from the rotating neutron star.

\begin{figure*}
\includegraphics[width=\textwidth]{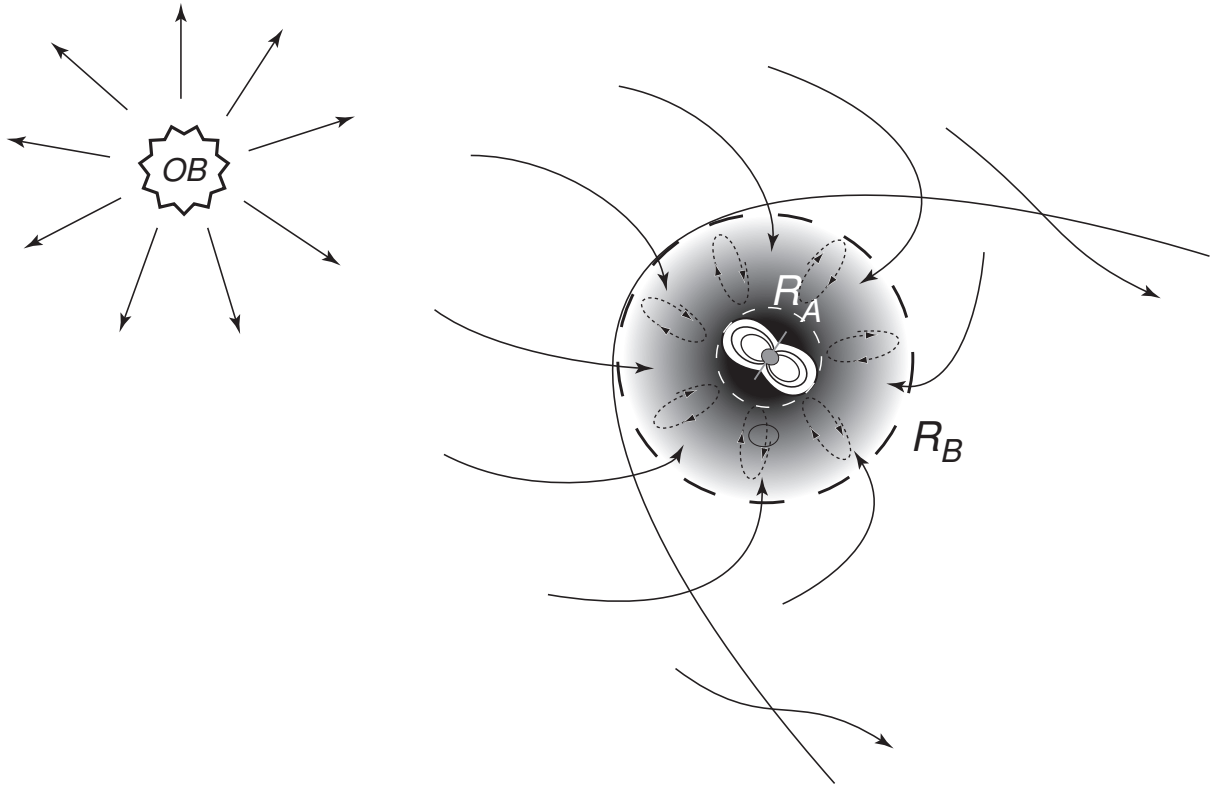}
\caption{A schematic picture of quasi-spherical accretion from the stellar wind of an optical companion star (left) onto a neutron star (right) in a binary system. In the regime of subsonic accretion, a quasi-spherical shell (shaded area) of radius $R_A$ is formed between the bow shock (parabolic curve) and the rotating magnetosphere. In this shell, large-scale convective motions are formed that may act to remove angular momentum from the magnetosphere. The outer radius of the shell is determined by the gravitational capture or Bondi radius  $R_B$. The characteristic velocity of the wind is $v_w\sim 300-1000$~km/s. The contour arrow shows the direction of the orbital velocity of the neutron star $v_{orb}$.}
\label{f:schema}
 \end{figure*}

There are several models in the literature (see especially \cite{IllarionovKompaneets90} and \cite{BisnovatyiKogan91}), from which the expression for the spin-down torque 
for quasi-spherically accreting neutron stars in the form 
$K_{sd}\sim -\dot M R_A^2 \omega^*$ can be derived.
Using the standard expression for the Alfv\'en radius, $R_A\sim \dot M^{-2/7}\mu^{4/7}$ this torque is proportional to $K_{sd}\sim -\mu^{8/7}\dot M^{3/7}$.
In our model, the matter in the shell settles subsonically as the region close to the magnetospheric surface cools down, and the Alfv\'en radius has a different dependence on the mass accretion rate and the magnetic field, $R_A\sim \dot M^{-2/11}\mu^{6/11}$ (see below).

One can show that there are two different mechanisms through which angular momentum can be transferred through a quasi-spherical shell. In the first case (we call this case {\textit moderate coupling}), angular momentum is transferred by convective motions in the shell. The breaking torque in the regime of settling accretion with convective removal of angular momentum depends on the accretion rate as  $K_{sd}\sim -\dot M^{3/11}$ (see Section \ref{s:main}). The velocity of the convective motions in this regime is close to the sound speed. It is also possible to have a settling regime where the angular momentum is removed by shear turbulence in the shell (the weak coupling regime). In this regime the characteristic velocities of the shear flow close to the magnetosphere is of the order of the linear rotational velocity. In this case $K_{sd}\sim \mu^2/R_c^3\sim \mu^2\omega^{*2}/(GM)$, i.e. in the weak coupling regime the torque does not depend on the accretion rate at all.

To stress the difference between the two possible regimes of subsonic accretion (with moderate and weak coupling), we can rewrite the expression for the breaking torque with convection (moderate coupling) using the corotational radius  and the Alfv\'en radius in the form $K_{sd}\sim -\mu^2/\sqrt{R_c^3 R_A^3}\sim -(\mu^2/R_c^3)(R_c/R_A)^{3/2}$ (see further details in Section 3). Since the factor $(R_c/R_A)^{3/2}\sim(\omega_K(R_A)/\omega^*)$ can be of the order of 10 or more in real systems, using a braking torque in the form of $\mu^2/R_c^3$ may lead to a strong overestimate
of the magnetic field of the neutron star.

The dependence of the braking torque on the accretion rate in the case 
of quasi-spherical settling accretion
suggests that variations of the mass accretion rate (and X-ray luminosity) 
must lead to a transition from spin-up (at high accretion rates) to spin-down (at small accretion rates) at some critical
value of $\dot M$ (or $R_A$), that differs from source to source. This phenomenon 
(known as torque reversal) 
is actually observed in wind-fed pulsars like Vela X-1, GX 301-2 and GX 1+4,
which we shall consider below in more detail.

The structure of this paper is as follows. In Section 2, we present an outline of the theory for quasi-spherical accretion onto a neutron star magnetosphere. We show that it is possible to construct a hot envelope around the neutron star through which subsonic accretion can take place and act to either spin up or spin down the neutron star. In Section 3, we discuss the structure of the interchange instability region which determines whether the plasma can enter the magnetosphere of the rotating neutron star. In Section 4 we consider how the spin-up/spin-down torques vary with a changing accretion rate. In Section 5, we show how to determine the parameters of quasi-spherical accretion from observational data. In Section 6, we apply our methods to the specific pulsars GX 301-2, Vela X-1, GX 1+4, SXP 1062 and 4U 2206+54. In Section 7 we discuss our results and, finally, in Section 8 we present our conclusions. A detailed gas-dynamic
treatment of the problem is presented in five appendices, which are very important to understand the physical processes involved. 

This work follows to a large extent the earlier published paper of \cite{Shakura_ea12}. However, here are included several additions, clarifying and refining the physical model (especially in Sections 1-4 and in Conclusions.

\section{Quasi-spherical accretion}
\label{s_qsaccr}
 
\subsection{The structure of a subsonic shell around a neutron star magnetosphere}
\label{s:shell} 
We shall here consider the torques applied to a neutron star in the case of quasi-spherical accretion from a stellar wind.
 Wind matter is gravitationally captured by the moving neutron star and a bow-shock is formed at a characteristic 
distance $R\sim R_B$, where $R_B$ is the Bondi radius. Angular momentum can be removed from the neutron star 
magnetosphere in two ways --- either with matter expelled from the magnetospheric boundary without accretion (the propeller regime, \cite{IllarionovSunyaev75}), or via large-scale convective motions
in a subsonic quasi-static shell around the magnetosphere, in which case the accretion rate onto the neutron star is determined by the ability of the plasma to enter the magnetosphere, in the regime of subsonic accretion.

In such a quasi-static shell, the temperature will be high (of the order of the virial temperature, see \cite{DaviesPringle81}), and the important point is whether hot matter from the shell can in fact enter the magnetosphere.  Two-dimensional calculations by \cite{ElsnerLamb77} have shown that hot monoatomic ideal plasma is stable relative to the Rayleigh-Taylor instability at the magnetospheric boundary, and plasma cooling is thus needed for accretion to begin. However, a closer inspection of the 3-dimensional calculations by \cite{AronsLea76a} reveals that the hot plasma is only marginally stable at the magnetospheric equator (to within 5\% accuracy of their calculations). Compton cooling and the possible presence of dissipative phenomena (magnetic reconnection etc.) facilitates the plasma entering the magnetosphere. 
We will show below that spin-down of the neutron star is possible in the case of accretion of matter from a hot envelope in the subsonic settling regime. 

To a zeroth approximation, we can neglect both rotation and radial motion (accretion) of matter
in the shell and consider only its equilibrium hydrostatic structure.
The radial velocity of matter falling through the shell $u_R$ is   
lower than the sound velocity $c_s$. Under these assumptions,  
the characteristic cooling/heating time-scale 
is much larger than the free-fall time-scale. 

In the general case where both gas pressure and anisotropic turbulent motions are present, 
Pascal's law is violated.  Then the hydrostatic equilibrium equation can be derived 
from the equation of motion \Eq{v_R1} with stress tensor components \Eq{W_RR} - \Eq{W_pp}
and zero viscosity (see Appendix A for more detail): 
\beq{e1}
-\frac{1}{\rho}\frac{dP_g}{dR}-
\frac{1}{\rho R^2}\frac{d(P_\parallel^t R^2)}{dR}+\frac{2P_\perp^t}{\rho R}-\frac{GM}{R^2}=0
\eeq 
Here $P_g=\rho c_s^2/\gamma$ is the gas pressure, and $P^t$ stands for the pressure due to turbulent 
motions:
\beq{ppar}
P_\parallel^t =\rho <u_\parallel^2>=\rho m_\parallel^2 c_s^2=\gamma P_g  m_\parallel^2
\eeq
\beq{pperp}
P_\perp^t =\rho <u_\perp^2>=\rho m_\perp^2 c_s^2 =\gamma P_g  m_\perp^2
\eeq 
($<u_t^2>=<u_\parallel^2>+2<u_\perp^2>$ is the  
turbulent velocity dispersion, $m_\parallel^2$ and $m_\perp^2$ are 
turbulent Mach numbers squared in the radial and tangential directions, respectively;
for example, in the case of isotropic turbulence $m_\parallel^2=m_\perp^2=(1/3)m_t^2$ where $m_t$ is the turbulent Mach number).
The total pressure is the sum of the gas and turbulence terms: $P_g+P_t=P_g(1+\gamma m_t^2)$.

The turbulent Mach number in the shell may in general depend on the radius. In our case, however, we will consider it constant. Furthermore, in real pulsars, turbulent heating (important from a dynamic point of view, see Appendix E) will change the estimated parameters by less than a factor of 2 (see formulas in Section 6).

We shall consider, to a first approximation, that the entropy $S$ is constant throughout the shell. For an ideal gas with adiabatic index $\gamma$ and equation of state $P=Ke^{S/c_V}\rho^\gamma$, the density can be expressed as a function of temperature: $\rho \sim T^{1/(\gamma-1)}$. Integrating the hydrostatic
equilibrium equation \Eq{e1}, we find:
\beq{hse_sol}
\frac{{\cal R} T}{\mu_m} = \myfrac{\gamma-1}{\gamma}\frac{GM}{R}\myfrac{1}
{1+\gamma m_\parallel^2-2(\gamma-1)(m_\parallel^2-m^2_\perp)}=\frac{\gamma-1}{\gamma}\frac{GM}{R}\psi(\gamma, m_t)\,.
\eeq 
(In this solution we have neglected the integration constant, which is not important 
deep inside the shell. It is important in the outer part of the shell, but since the outer region close to the bow shock at $\sim R_B$ is not spherically 
symmetric, its structure can only be found numerically). 
We note that taking turbulence into account somewhat decreases the temperature within the shell. Most important, however, is that the anisotropy of turbulent motions, caused
by convection, in the stationary case changes the distribution of angular velocity 
in the shell. Below we will show that in the case of isotropic turbulence, the 
angular velocity distribution within the shell is close to quasi-Keplerian: $\omega(R) \sim R^{-3/2}$. In the case of strongly anisotropic turbulence caused by convection, $m_\parallel^2\gg m_\perp^2$, the distribution of momentum in the shell may become almost iso-angular: $\omega(R) \sim R^{-2}$. Below we shall see that an analysis 
of several real X-ray pulsars favors an iso-angular momentum rotation distribution. 

Now, let us write down how the density varies inside the quasi-static shell for $R\ll R_B$.
For a fully ionized gas with $\gamma=5/3$ we find:
\beq{rho(R)}
\rho(R)=\rho(R_A)\myfrac{R_A}{R}^{3/2}
\eeq

\beq{P(R)}
P(R)=P(R_A) \myfrac{R_A}{R}^{5/2}\,.
\eeq 
The above equations describe the structure of an ideal static adiabatic shell 
above the magnetosphere. Of course, at $R\sim R_B$ the problem is essentially non-spherically symmetric and
numerical simulations are required. 

Corrections to the adiabatic temperature gradient due to 
convective energy transport through the shell are calculated in Appendix D.

\subsection{The Alfv\'en surface}
At the magnetospheric boundary (the Alfv\'en surface), the total pressure (including
isotropic gas pressure and the possibly anisotropic turbulent pressure) is balanced by 
the magnetic pressure $B^2/(8\pi)$
\beq{}
P_g+P_t=P_g(R_A)(1+\gamma m_t^2)=\frac{B^2(R_A)}{8\pi}\,.
\eeq 
The magnetic field at the Alfv\'en radius is determined by 
the dipole moment and magnetic field of the neutron star and by electric currents flowing on the Alfv\'enic surface (in the magnetopause):
\beq{P(RA)}
P_g(R_A)=\frac{K_2}{(1+\gamma m_t^2)}\frac{B_0^2}{8\pi} \myfrac{R_0}{R_A}^6 =\frac{\rho{\cal R}T}{\mu_m}
\eeq 
where the dimensionless coefficient $K_2$ takes into account the contribution from these currents 
and the factor $1/(1+\gamma m_t^2)$ is due to the turbulent pressure term. For example, in the model by Arons and Lea \cite{AronsLea76a} (their Eq. 31), $K_2=(2.75)^2\approx 7.56$. At the magnetospheric cusp (where the magnetic force line is branched),
the radius of the Alfv\'en surface is about 0.51 times that of the equatorial radius \cite{AronsLea76a}. Below we shall assume that $R_A$ is the equatorial radius 
of the magnetosphere, unless stated otherwise. 

The plasma is able to enter the magnetosphere mainly due to the interchange instability. 
In the stationary regime, let us introduce the accretion rate $\dot M$ onto the neutron star surface. 
From the continuity equation in the shell we find
\beq{rho_cont}
\rho(R_A)=\frac{\dot M}{4\pi u_R(R_A) R_A^2}
\eeq
Clearly, the velocity of absorption of matter by the magnetosphere is smaller than 
the free-fall velocity, so we introduce a dimensionless factor $f(u)=u_R/\sqrt{2GM/R}<1$.
Then the density at the magnetospheric boundary is
\beq{rho(R)}
\rho(R_A)=\frac{\dot M}{4\pi f(u) \sqrt{2GM/R_A} R_A^2}\,.
\eeq
For example, in the model calculations by \cite{AronsLea76a}, $f(u)\approx 0.1$;
in our case, at high X-ray luminosities, the value of $f(u)$ may attain $\approx 0.5$.  
If we imagine that the shell is impenetrable and that there is no accretion through it,  
$\dot M \to 0$. In this case $u_R\to 0$, $f(u)\to 0$, while the density 
in the shell remains finite. 
In some sense, the matter is leaking from the magnetosphere down onto the neutron star, 
and the leakage may be either very small ($\dot M \to 0$) or have a finite non-zero value ($\dot M \ne 0$). 

Plugging $\rho(R)$ into \Eq{P(RA)} and using \Eq{hse_sol} and 
the definition of the dipole magnetic moment
\[
\mu=\frac{1}{2}B_0R_0^3
\] 
(where $R_0$ is the neutron star radius), we find an expression for the Alfv\'en radius in the case of quasi-spherical accretion:
\beq{RA_def}
R_A=\left[\frac{4\gamma}{(\gamma-1)}\frac{f(u) K_2}{\psi(\gamma, m_t)(1+\gamma m_t^2)} \frac{\mu^2}{\dot M\sqrt{2GM}}\right]^{2/7}\,.
\eeq 

It should be stressed that in the presence of a hot shell the Alfv\'en radius is
determined by the static gas pressure (with a possible addition of turbulent motions) at the magnetospheric boundary, 
which is non-zero even for a zero-mass accretion rate through the shell. The dependence of $f(u)$ on the accretion rate $\dot M$ in the case of a settling shell taking cooling into account will be derived below (see \Eq{fu} below). In the supersonic (Bondi) regime we obviously have $f(u)=1$. We note that accretion with subsonic velocity can take place even in the Bondi regime, but with significantly lower accretion rate (as compared to the maximum). In the Bondi regime (i.e. in the adiabatic regime without gas heating and/or cooling), the choice of solution depends on the boundary conditions. 

\subsection{The mean velocity of matter entering through the magnetospheric boundary}
\label{s:f(u)} 

As mentioned above, the plasma enters the magnetosphere of the slowly rotating neutron star due to 
the interchange instability. The boundary between the plasma and the magnetosphere is stable 
at high temperatures $T>T_{cr}$, but becomes unstable at $T<T_{cr}$, and remains in 
a neutral equilibrium at $T=T_{cr}$ \cite{ElsnerLamb77}. The critical temperature is:
\beq{Tcr}
{\cal R}T_{cr}=\frac{1}{2(1+\gamma m_t^2)}\frac{\cos\chi}{\kappa R_A}\frac{\mu_mGM}{R_A}\,.
\eeq 
Here $\kappa$ is the local curvature of the magnetosphere, $\chi$ is the angle 
the outer normal makes with the radius-vector at a given point, and the contribution 
of turbulent pulsations in the plasma to the total pressure is
taken into account by the factor $(1+\gamma m_t^2)$.
The effective gravity acceleration can be written as
\beq{g_eff}
g_{eff}=\frac{GM}{R_A^2}\cos\chi\left(1-\frac{T}{T_{cr}}\right)\,.
\eeq 
The temperature in the quasi-static shell is given by \Eq{hse_sol}, and 
the condition for the magnetosphere instability can thus be rewritten as:
\beq{m_inst}
\frac{T}{T_{cr}}=\frac{2(\gamma-1)(1+\gamma m_t^2)}
{\gamma}\psi(\gamma,
 m_t) \frac{\kappa R_A}{\cos\chi}<1\,.
\eeq 

According to \cite{AronsLea76a}, when the external gas pressure decreases 
with radius as $P\sim R^{-5/2}$, the form of the magnetosphere far from the 
polar cusp can be described to within 10\% accuracy as $(\cos\lambda)^{0.2693}$
(here $\lambda$ is the polar angle counting from the magnetospheric equator). The instability first appears near the equator, where the curvature is minimal. Near the equatorial plane ($\lambda=0$),
for a poloidal dependence of the magnetosphere $\approx (\cos \lambda)^{0.27}$ 
we get for the curvature $k_pR_A=1+0.27$. The toroidal field curvature at the magnetospheric equator is  $k_t=1$. The tangent sphere at the equator cannot have a radius larger than the inverse poloidal curvature, therefrom $\kappa R_A=1.27$ at $\lambda=0$. This is somewhat larger than the value of $\kappa R_A=\gamma/(2(\gamma-1))=5/4=1.25$ (
for $\gamma=5/3$ in the absence of turbulence or
for fully isotropic turbulence),
but within the accuracy limit\footnote{In \cite{AronsLea76b}, the curvature is calculated to be $\kappa R_A\approx 1.34$, still within the accuracy limit}. The contribution from anisotropic turbulence 
decreases the critical temperature; for example,
for $\gamma=5/3$, in the case of strongly anisotropic turbulence 
$m_\parallel=1$, $m_\perp=0$, at $\lambda=0$ we obtain $T/T_{cr}\sim 2$, 
i.e. anisotropic turbulence
increases the stability of the magnetosphere.
So initially the plasma-magnetospheric boundary
is stable, and after cooling to $T<T_{cr}$ the plasma instability sets in, starting in the equatorial zone, where the curvature of the magnetospheric surface is minimal.  

Let us consider the development of the interchange instability when cooling 
(predominantly Compton cooling) is present. The temperature changes as \cite{Kompaneets56}, \cite{Weymann65}
\beq{dTdt}
\frac{dT}{dt}=-\frac{T-T_x}{t_C}\,,
\eeq 
\beq{t_comp}
t_{C}=\frac{3}{2\mu_m}\frac{\pi R_A^2 m_e c^2}{\sigma_T L_x}
\approx 10.6  [\hbox{s}] R_{9}^2 \dot M_{16}^{-1}\,.
\eeq 
where the Compton cooling time is
\beq{t_comp}
t_{C}=\frac{3}{2\mu_m}\frac{\pi R_A^2 m_e c^2}{\sigma_T L_x}
\approx 10.6  [\hbox{s}] R_{9}^2 \dot M_{16}^{-1}\,.
\eeq
Here $m_e$ is the electron mass, $\sigma_T$ is the Thomson cross section, $L_x=0.1 \dot M c^2$ is the X-ray luminosity, $T$ is the electron temperature (which is equal to the ion temperature since the timescale of electron-ion energy exchange here is the shortest possible), $T_x$ is the X-ray temperature and
$\mu_m=0.6$ is the molecular weight. The photon temperature is $T_x=(1/4) T_{cut}$ for a bremsstrahlung spectrum with an exponential cut-off at $T_{cut}$, typically $T_x=3-5$~keV. 
The solution of equation \Eq{dTdt} reads:
\beq{}
T=T_x+(T_{cr}-T_x)e^{-t/t_C}\,.
\eeq 
We note that $T_{cr}\sim 30\,\hbox{keV}\gg T_x\sim 3$~keV. It is seen that for $t\approx 2t_C$ the temperature decreases to $T_x$. In the linear approximation the temperature changes as:
\beq{tlin}
T\approx T_{cr}(1-t/t_C)\,.
\eeq 
Plugging this expression into \Eq{g_eff}, we find that the effective gravity acceleration increases linearly with time as:
\beq{}
g_{eff}\approx \frac{GM}{R_A^2}\frac{t}{t_C}\cos\chi \,.
\eeq 
Correspondingly, the velocity of matter due to the instability growth increases with time as:
\beq{u}
u_r=\int\limits_0^{t_{inst}} g_{eff} dt=\frac{GM}{R_A^2}\frac{t_{inst}^2}{2t_C}\cos\chi \,.
\eeq 
Here, $t_{inst}$ is the characteristic time of the instability which can be expressed in the form: 
\beq{tinst}
t_{inst}=\frac{K_0}{\omega_K(R_A)}\frac{u_{ff}}{u_r}=\frac{K_0}{\omega_K(R_A)f(u)}\,.
\eeq 
The choice of this expression is due to the fact that in the case of rapid cooling, the velocity of matter $u_r$ is of the order of the free-fall time $u_{ff}$, and for slow cooling $u_r\ll u_{ff}$. We have also defined $f(u)\equiv u_r/u_{ff}<1$, which will be used in the following. $K_0$ is a dimensionless constant of the order of unity.

Plugging $t_{inst}$ into \Eq{u}, we find the velocity obtained by the matter during the time-scale of the instability:
\beq{}
u_r(t_{inst})=\frac{K_0^2}{2}\frac{R_A}{t_C f(u)^2}\cos\chi\,.
\eeq
Dividing both parts of this equation by $u_{ff}$ and solving for $f(u)$,
we get the expression for $f(u)$: 
\beq{fu1}
f(u)=\myfrac{K_0^2}{2}^{1/3}\myfrac{t_{ff}}{t_C}^{1/3}(\cos\chi)^{1/3} .
\eeq 
We used here the expression for the free-fall time:
\beq{}
t_{ff}\equiv \frac{R_A}{u_{ff}(R_A)}=\frac{R_A^{3/2}}{\sqrt{2GM}}\,.
\eeq
Then, the characteristic time-scale for the instability can be rewritten in the form:
\beq{tinst1}
t_{inst}=\frac{(2K_0)^{1/3}}{\omega_K(R_A)}\myfrac{t_C}{t_{ff}}^{1/3}(\cos\chi)^{-1/3}.
\eeq 

From this it can be seen that for $t_C\gg t_{ff}$, the timescale for the instability is much larger than the free-fall time. 
\beq{tinst2}
\frac{t_{inst}}{t_{ff}}=2^{1/2}(2K_0)^{1/3}\myfrac{t_C}{t_{ff}}^{1/3}(\cos\chi)^{-1/3}
\eeq
On the other hand, the time-scale of the instability is shorter than the Compton cooling time:
\beq{tinst3}
\frac{t_{inst}}{t_C}=2^{1/2}(2K_0)^{1/3}\myfrac{t_{ff}}{t_{C}}^{2/3}(\cos\chi)^{-1/3}<1\,,
\eeq
which allows us to use the linear expansion of temperature increase 
as a function of time time \Eq{tlin}.

The characteristic scale of instability growth is:
\beq{Delta}
\Delta=\int\limits_0^{t_{inst}} u_rdt=\frac{1}{6}\frac{GM}{R_A^2}\frac{t_{inst}^3}{t_C}\cos\chi=\frac{1}{3}u_rt_{inst}=\frac{\sqrt{2}}{3}K_0R_A\,.
\eeq
In this way, during $t_{inst}$, the scale of the instability becomes comparable to the magnetospheric radius, and the settling velocity turns out to be 
much smaller than free-fall velocity $u_{ff}$. 
Clearly, later in the non-linear stage of the instability growth the velocity of matter 
approaches the
free-fall velocity. We mainly consider the linear stage, since 
at this stage the temperature is still high enough (although the entropy 
starts decreasing with decreasing radius), and it is in this zone that a toroidal component of the magnetic field is formed and effective angular 
momentum transfer from the magnetosphere to the shell can take place. At later stages of
instability growth, the loss of entropy is too strong for convection to begin. 

Let us estimate the accuracy of our approximation by retaining the second-order terms
in the exponent expansion. Then the velocity the matter acquires during 
the instability time $t_{inst}$ is: 
\beq{2dorder}
u_r(t_{inst})=K_0^{2/3}\myfrac{GM}{t_C}^{1/3}(\cos\chi)^{1/3}\left[1-\frac{2^{5/6}K_0^{1/3}}{3}\myfrac{t_{ff}}{t_{C}}^{2/3}(\cos\chi)^{-1/3}\right]\,.
\eeq
Clearly, the smaller accretion rate, the smaller the ratio $t_{ff}/t_C$, and the better our approximation. 

We note that for the magnetospheric radius in the form  $\sim \cos\lambda^n$ we have $\tan\chi=n\tan\lambda$. Therefore, for $n\simeq 0.27$ close to the equator $\cos\chi\simeq 1$ with high accuracy, and we will in the following ignore this factor. We also note that in the magnetospheric cusp region $\cos\chi\simeq 0$, and in this region matter can almost not enter the magnetosphere at all.
Substituting \Eq{t_comp} into \Eq{fu1} and then $f(u)$ into definition \Eq{RA_def}, we find for $\gamma=5/3$ the expression for the Alfv\'en radius in 
this regime:
\beq{RA}
R_A\approx 1.55\times 10^9[\hbox{cm}] K_0^{2/11}
[(1+\frac{5}{3} m_t^2)\psi(\frac{5}{3}, m_t)]^{-3/11}\myfrac{\mu_{30}^3}{\dot M_{16}}^{2/11}\,.
\eeq 
We stress the difference of the obtained expression for the Alfv\'en radius with the 
standard one, $R_A\sim \mu^{4/7}/\dot M^{-2/7}$, which is obtained by equating the dynamical
pressure of falling gas to the magnetic field pressure; this difference comes from the dependence of 
$f(u)$ on the magnetic moment and mass accretion rate in the settling accretion regime. 

The coefficient due to turbulence
\beq{tur}
K_t=(1+\frac{5}{3} m_t^2)\psi(\frac{5}{3}, m_t)
\eeq
is obviously equal to 1 for isotropic turbulence (see the expression for $\psi$ \Eq{hse_sol}), and thus of interest only in the case of anisotropic turbulence.

Plugging \Eq{RA} into \Eq{fu1}, we obtain an explicit expression for $f(u)$:
\beq{fu}
f(u)\approx 0.39 K_0^{7/11}K_t^{1/22}\dot M_{16}^{4/11}\mu_{30}^{-1/11}\,.
\eeq 

A necessary condition for removal of angular momentum from the magnetosphere via convection is the condition of subsonic settling (the Mach number for the settling velocity ${\cal M}\equiv u_r/c_s <1$), which for  $\gamma=5/3$ is reduced to the inequality $f(u)<1/\sqrt{3}$.
Clearly, this condition is fulfilled for mass accretion rates around $10^{16}$~g/s and lower.
It is also important to stress that convection in the shell as well as removal of angular momentum practically stops working when the mean radial settling velocity of the matter $u_r$ becomes higher than the convective velocity $u_c$, i.e. when the convective Mach number $m_c=u_c/c_s\sim m_t$ becomes smaller than the standard Mach number ${\cal M}=u_r/c_s$. And oppositely, when the Mach number of the radial flow becomes smaller than the turbulent Mach number ${\cal M}<m_t\sim m_c$, removal of angular momentum through the shell may take place.
When the accretion rate of matter through the shell becomes larger than a certain critical value $\dot M>\dot M^\dag$ the velocity of the accretion flow close to the Alfv\'enic surface may become higher than the sound speed, and a supersonic flow region with matter in free fall may form above the magnetosphere. Through this region it is not possible to remove any angular momentum from the rotating magnetosphere. In this case, settling accretion is not applicable. A shockwave forms above the magnetosphere and plasma interaction with the magnetosphere is described in the scenario studied in e.g. \cite{Burnard_ea83}. Depending on the inhomogeneity of the captured stellar wind, the specific angular momentum may be either positive or negative, and thus alternating episodes of spin-up and spin-down of the neutron star are possible in the supersonic regime. 
It is easy to estimate the critical X-ray luminosity above which the transition from the subsonic (at low X-ray luminosities) to the Bondi-Hoyle-Littleton (at high X-ray luminosities) regime takes place. Indeed, assuming a limit for the dimensionless settling velocity of $f(u)$=0.5 (at which removal of angular momentum through the shell is still possible, see further Appendix E), from equation \Eq{fu}, we find the maximum possible value of the accretion rate for the settling regime with removal of angular momentum:
\beq{M*}
\dot M^\dag_{16}\approx 2 K_0^{-7/4}K_t^{-1/8}\mu_{30}^{1/4}\,.
\eeq  
We note that a similar value for the critical accretion rate can be found from a comparison of the Compton cooling time to the time-scale for convection close to the Alfv\'en radius.

To conclude this section, we note that it is not difficult to perform a similar analysis for the velocity of matter in the magnetosphere due to radiative cooling of the plasma, for cases when Compton cooling is less effective \cite{Shakura_ea12b}. This scenario may be realized in X-ray pulsars at very low accretion rates when the shape of the X-ray beam-pattern changes and the photon beam forms a pencil diagram illuminating the magnetospheric cusp. 
In this way one can explain the episodic <<off-states>> (with very low X-ray luminosity), accompanied with a phase-shift in the X-ray pulse profile \cite{Doroshenko_ea11} as observed in pulsars like e.g. Vela X-1.

\section{Transfer of angular momentum to the magnetosphere}
\label{s:angmom}
Let us now consider a quasi-stationary subsonic shell in which accretion proceeds 
onto the neutron star magnetosphere. We stress that in this regime, i.e. the settling regime, the accretion rate onto the neutron star is determined by the density at the bottom of the shell (which is directly related to 
the density downstream the bow shock in the gravitational capture region) and the ability of the plasma to enter the magnetosphere through the Alfv\'enic surface.

The rotation law in the shell depends on the treatment of the turbulent viscosity
(see Appendix B for cases when the Prandtl law and isotropic turbulence are applicable) and 
the possible anisotropy of the turbulence due to convection (see Appendix C). In the latter case the anisotropy leads to more powerful radial turbulence than perpendicular. In this way, as shown in Appendix B and C,
we arrive at a set of quasi-power-law solutions for the radial dependence of the angular 
rotation velocity in a convective shell. We shall in the following consider a simple power-law dependence of the angular momentum on radius,
\beq{rotation_law}
\omega(R)\sim R^{-n}.
\eeq
In Section \ref{s:applications} in applications to real pulsars we will use a quasi-Keplerian law with $n=3/2$ as well as an iso-angular momentum distribution with $n=2$, which in some sense represent limiting cases
among possible solutions.

When approaching the bow shock, $R \to R_B$, and the angular velocity of matter approaches the orbital velocity, $\omega\to \omega_B$. Close to the bow shock the problem is not spherically
symmetric any more since the flow becomes very complex (parts of the flow may cause the hot shell to bend, etc.),
and the structure of the flow can be studied only using numerical simulations. In the absence of such simulations, we shall assume that the assumption of an iso-angular momentum distribution is valid up to the 
front of the bow shock located at a distance from the neutron star which we shall take to be the Bondi radius $R_B$, 
\[
R_B\simeq 2GM/(V_w^2+v_{orb}^2)^2
\]
where $V_w$ is the stellar wind velocity at the neutron star orbital distance, and $v_{orb}$ is the neutron star orbital velocity.

This means that the angular velocity of rotation of matter near the magnetosphere $\omega_m$
will be related to $\omega_B$ via
\beq{omega_m1}
\omega_m= \tilde\omega\omega_B\myfrac{R_B}{R_A}^{n}.
\eeq
(Here the numerical factor $\tilde\omega>1$ takes into account the deviation 
of the actual rotational law from the value obtained by using the assumed power-law dependence near the Alfv\'en radius; see Appendix B and C for more detail.)

Now, let the NS magnetosphere rotate with 
an angular velocity $\omega^*=2\pi/P^*$ where $P^*$ is the neutron star spin period. 
The matter at the bottom of the shell 
rotates with an angular velocity $\omega_m$, in general different from $\omega^*$.
If $\omega^*>\omega_m$, coupling of the plasma with the magnetosphere ensures transfer of angular momentum from the magnetosphere to the shell, or from the shell to the
magnetosphere if $\omega^*<\omega_m$. In the general case, the coupling of matter with the magnetosphere can be moderate or strong. In the strong coupling regime the toroidal magnetic field component $B_t$ is proportional to the 
poloidal field component $B_p$ as $B_t\sim -B_p (\omega_m-\omega^*)t$, and $|B_t|$ can grow to $\sim |B_p|$. This regime can be expected for rapidly rotating magnetospheres 
when $\omega^*$ is comparable to or even greater than the Keplerian
angular frequency $\omega_K(R_A)$; in the latter case the propeller regime sets in. 
In the moderate coupling regime, the plasma can enter the magnetosphere due to instabilities on a timescale shorter than the time needed for the toroidal field to grow to the value of the poloidal field, 
so $B_t < B_p$.  

\subsection{The case of strong coupling}
\label{s:strongcoupling}
Let us first consider the strong coupling regime. In this regime, powerful large-scale convective motions may lead to turbulent magnetic field diffusion accompanied by magnetic field dissipation. 
This process is characterized by the turbulent magnetic field diffusion coefficient $\eta_t$.  In this case 
the toroidal magnetic field (see e.g. \cite{Lovelace_ea95} and references therein) is:
\beq{bt}
B_t=\frac{R^2}{\eta_t}(\omega_m-\omega^*)B_p\,.
\eeq
The turbulent magnetic diffusion coefficient is related to the kinematic turbulent viscosity as
$\eta_t\simeq \nu_t$. The latter can be written as:
\beq{nut}
\nu_t=<u_tl_t>\,.
\eeq  

According to the phenomenological Prandtl law, the average characteristics of 
a turbulent flow (the velocity $u_t$, the characteristic scale of turbulence $l_t$ 
and the shear $\omega_m-\omega^*$) are related as:
\beq{Prandtl}
u_t\simeq l_t |\omega_m-\omega^*|\,.
\eeq
In our case, the turbulent scale must be determined by the largest scale of energy supply to the turbulence 
from the rotation of the non-spherical magnetospheric surface. This scale is determined by the difference in velocity between the solidly rotating magnetosphere and the accreting matter that is still not interacting with the magnetosphere, i.e. $l_t\simeq R_A$, which determines the turn-over velocity of the largest turbulence eddies. At smaller scales a turbulent cascade develops. Substituting this scale into equations \Eq{bt}-\Eq{Prandtl} above, we find that in the strong coupling regime $B_t\simeq B_p$.

The momentum of the forces due to plasma-magnetosphere interactions acts on the neutron star and 
changes its spin according to: 
\beq{17}
I\dot \omega^*=\int\frac{B_tB_p}{4\pi}\varpi dS = 
\pm \tilde K(\theta)K_2\frac{\mu^2}{R_A^3}
\eeq
where $I$ is the neutron star's moment of inertia, $\varpi$ is the distance from the rotational axis and $\tilde K(\theta)$ is a numerical coefficient depending
on the angle between the rotational and magnetic dipole axes. The coefficient
$K_2$ appears in the above expression for the same reason as in \Eq{P(RA)}.
The positive sign corresponds to positive flux of angular momentum to the neutron star 
($\omega_m>\omega^*$). The negative sign corresponds to negative flux of angular momentum across 
the magnetosphere ($\omega_m<\omega^*$). 

At the Alfv\'en radius, the matter couples with the magnetosphere and acquires the angular velocity of the neutron star. It then falls onto the neutron star surface
and returns the angular momentum acquired at $R_A$ back to the neutron star via the magnetic field. As a result of this process, the neutron star spins up at a rate determined by the expression:
\beq{suz}
I\dot \omega^*=+z \dot M R_A^2\omega^*
\eeq
where $z$ is a numerical coefficient which takes into account the angular momentum of the falling matter. If all matter falls from the equatorial equator, $z=1$; if matter falls strictly along the spin axis, $z=0$. 
If all matter were to fall across the entire magnetospheric surface, then $z=2/3$. 

Ultimately, the total torque applied to the neutron star in the strong coupling regime yields
\beq{sd_eq_strong}
I\dot \omega^*=\pm \tilde K(\theta)K_2 \frac{\mu^2}{R_A^3} +z \dot M R_A^2\omega^*\,.
\eeq 
Using \Eq{RA_def}, we can eliminate $\dot M$ in the above equation to obtain in the spin-up regime ($\omega_m>\omega^*$)
\beq{su}
I\dot \omega^*=\frac{\tilde K(\theta)K_2\mu^2}{R_A^3}
\left[1+z\frac{4\gamma f(u)}{\sqrt{2}(\gamma-1)(1+\gamma m_t^2)\psi(\gamma, m_t)\tilde K(\theta)}\myfrac{R_A}{R_c}^{3/2}\right]
\eeq
where $R_c^3=GM/(\omega^*)^2$ is the corotation radius. In the spin-down regime ($\omega_m<\omega^*$) we find
\beq{sd}
I\dot \omega^*=-\frac{\tilde K(\theta)K_2\mu^2}{R_A^3}
\left[1-z\frac{4\gamma f(u)}{\sqrt{2}(\gamma-1)(1+\gamma m_t^2)\psi(\gamma, m_t)\tilde K(\theta)}\myfrac{R_A}{R_c}^{3/2}\right]\,.
\eeq
Note that in both cases $R_A$ must be smaller than $R_c$, otherwise the propeller effect prohibits accretion. In the propeller regime $R_A>R_c$, 
matter does not fall onto the neutron star, there are no accretion-generated 
X-rays from the neutron star, the shell rapidly cools down and
shrinks and the standard Illarionov and Sunyaev propeller regime \cite{IllarionovSunyaev75}, with matter outflow from the magnetosphere, is established.

During both spin-up and spin-down, 
the neutron star angular velocity $\omega^*$ almost approaches the angular velocity of
matter at the magnetospheric boundary, $\omega^*\to \omega_m(R_A)$. The difference between $\omega^*$ and
$\omega_m$ is small so the second term in the square brackets in \Eq{su} and \Eq{sd} is much smaller than unity. Also note that when approaching the propeller regime ($R_A\to R_c$), the accretion rate decreases, 
$f(u)\to 0$, the second term in the square brackets vanishes, and the spin evolution is determined solely by the spin-down term $-\tilde K(\theta) \mu^2/R_A^3$. (In the
propeller regime, $\omega_m< \omega_K(R_A)$, $\omega_m<\omega^*$, $\omega^*> \omega_K(R_A)$ ). So the neutron star spins down to the Keplerian frequency at the Alfv\'en radius.
In this regime, the specific angular momentum of the matter that flows in and out
from the magnetosphere is, of course, conserved. 

Near equilibrium ($\omega^*\sim \omega_m$),
relatively small fluctuations in $\dot M$ across the shell will lead to very strong fluctuations in 
$\dot \omega^*$ since the toroidal field component can change its sign by changing from $+B_p$ to $-B_p$. If strong coupling actually occurs in nature, this property would be a distinguishing feature of this regime. It is known (see eg. \cite{Bildsten_ea97}, \cite{Finger_ea11}) that 
real X-ray pulsars sometimes exhibit rapid spin-up/spin-down transitions not associated
with X-ray luminosity changes, which may be evidence that 
they temporarily enter the strong coupling regime. It can not be excluded that the
triggering of the strong coupling regime may be due to the magnetic field
frozen into the accreting plasma that has not yet entered the magnetosphere. Accretion of magnetized plasma onto neutron stars is studied in detail in the recent work by \cite{Ikhsanov12}. 

\subsection{The case of moderate coupling}
\label{s:moderatecoupling}

The strong coupling regime considered above may be realized in the extreme case where 
the toroidal magnetic field $B_t$ attains a maximum possible value  
$\sim B_p$ due to magnetic turbulent diffusion. Usually, the coupling of matter with the magnetosphere is mediated by 
different plasma instabilities whose characteristic times are too short for substantial toroidal field growth. As discussed above in Section \ref{s:shell}, the shell is very hot close to the magnetosphere boundary, so without cooling above it the plasma is marginally stable with respect to the interchange instability (according to the calculations by  \cite{AronsLea76a}). 

Let us write down the torque due to magnetic forces applied to the neutron star:
\beq{torquem}
I\dot \omega^*=\int\frac{B_tB_p}{4\pi}\varpi dS 
\eeq
On the other hand, there is a mechanical torque on the magnetosphere from the base of the shell caused by the turbulent stresses $W_{R\phi}$:
\beq{torquet}
\int W_{R\phi} \varpi dS\,,
\eeq
where the viscous turbulent stresses can be written as (see the Appendices for more details)
\beq{}
W_{R\phi}=\rho \nu_t R \frac{\partial \omega}{\partial R}\,.
\eeq
To scpecify the turbulent viscosity coefficient
\beq{}
\nu_t=\langle u_c l_t\rangle\,,
\eeq 
we assume that the characteristic scale of the turbulence close to the magnetosphere is $l_t\sim R_A$, and that the characteristic velocity of the turbulent pulsations is determined by the mechanism of turbulence in the plasma above the magnetosphere.
If there are strong convective motions in the shell, caused by heating of its base, then $u_c\sim c_s$, where $c_s$ is the sound speed. If convection is prohibited, there is still turbulence, caused by the shear  flow in the shell ($\omega \sim 1/R^2$, see the Appendices). In this case $u_c(R_A) \sim u_\phi(R_A)\sim \omega^*R_A\ll c_s$. Obviously, the ratio of the stresses for the different cases turns out to be of the order of $\omega^*/\omega_K(R_A)$, which for slowly rotating pulsars is around $0.03-0.3$. Equating the torques \Eq{torquem} and \Eq{torquet}, we get
\beq{}
\rho u_c R_A \frac{\partial \omega}{\partial R}=\frac{B_tB_p}{4\pi}
\eeq
We eliminate the density from this expression using the pressure balance at the magnetospheric boundary \Eq{P(RA)} and the expression for the temperature \Eq{hse_sol}, and make the substitution 
\beq{zeta}
\frac{\partial \omega}{\partial R}=\frac{\omega_m-\omega^*}{\zeta R_A}.
\eeq 
Here we have introduced the dimensionless factor $\zeta<1$, characterizing the size of the zone in which there is an effective exchange of angular momentum between the magnetosphere and the base of the shell. Then we find the relation between the toroidal and poloidal components of the magnetic field in the magnetosphere: 
\beq{btbp}
\frac{B_t}{B_p}=\frac{\gamma}{\sqrt{2}(\gamma-1)K_t}\myfrac{u_c}{u_{ff}}\myfrac{\omega_m-\omega^*}{\zeta \omega_K(R_A)}
\eeq
(Here and below we have used the following designations: the free fall velocity $u_{ff}\equiv \sqrt{\frac{2GM}{R}}$, the Keplerian frequency at the magnetospheric boundary  $\omega_K(R_A)$ and the correction coefficient due to turbulence $K_t\equiv (1+\gamma m_t^2)\psi(\gamma,m_t)$).

Substituting \Eq{btbp} into \Eq{torquem}, in case of convection $u_c=m_c c_s$ (where we have introduced the Mach number for convective motions $m_c$), the spin-down rate of the neutron star can be written as: 
\beq{sd1}
I\dot\omega^*=\myfrac{K_1}{\zeta}K_2\frac{\mu^2}{R_A^3}\frac{\omega_m-\omega^*}{\omega_K(R_A)}\,.
\eeq
where $K_1$ is a constant of the order of unity arising from a combination of the parameters in \Eq{btbp}.
In this case \Eq{btbp} can be re-written in the form
\beq{btbpK1} 
\frac{B_t}{B_p}=\tilde K\myfrac{K_1}{\zeta}\frac{\omega_m-\omega^*}{ \omega_K(R_A)}\,,
\eeq
where the geometrical factors arising from the integration of 
\Eq{torquem} are included in the coefficient $\tilde K\sim 1$. 

If the differential rotation at the base of the shell gives rise to turbulence, $u_c\sim u_\phi=\omega^*R_A$, and the expression for spin down takes the form 
\beq{sd1a}
I\dot\omega^*=\myfrac{\tilde K_1}{\zeta}K_2\frac{\mu^2}{R_A^3}\myfrac{R_A}{R_c}^{3/2}\frac{\omega_m-\omega^*}{\omega_K(R_A)}\,.
\eeq
where
\beq{Rcor}
R_c\equiv \myfrac{GM}{\omega^{*2}}^{1/3}
\eeq
is the corotational radius (see also \cite{Lipunov87}).

Evidently, the breaking torque is in this case smaller by a factor of $(R_A/R_c)^{3/2}$ as compared to when there are convective motions in the shell. We will call this case the case of  \textit{weak cuopling}. It can easily be seen that in this case both the breaking torque and the spin down rate of the neutron star are independent of the mass accretion rate (in the limit  $\omega_m\to 0$ we have just $K_{sd}\sim \mu^2/R_c^3$, \cite{Lipunov87}). As will be discussed later on, the non-equilibrium pulsar GX 1+4 shows during spin-down a negative correlation $\dot\omega^*$ with luminosity \cite{Chakrabarty_ea97}. Therefore, we prefer breaking according to \Eq{sd1} (i.e. with moderate coupling).

Using the definition of the Alfv\'en radius $R_A$ \Eq{RA_def} and the expression for the Keplerian frequency $\omega_K$, we can write \Eq{sd1} in the form
\beq{sd_om}
I\dot \omega^*=Z \dot M R_A^2(\omega_m-\omega^*).
\eeq
Here the dimensionless coefficient $Z$ is 
\beq{Zdef}
Z=\frac{\myfrac{K_1}{\zeta}}{f(u)}\frac{\sqrt{2}(\gamma-1)}{4\gamma}K_t\,.
\eeq
Substituting in this formula $\gamma=5/3$ and the expression  \Eq{fu1}, we find 
\beq{Znum}
Z\approx 0.363 \myfrac{K_1}{\zeta} K_0^{-7/11}K_t^{21/22} \dot M_{16}^{-4/11}\mu_{30}^{1/11}.
\eeq  

Taking into account that the matter that falls onto the neutron star adds the angular momentum
$z\dot M R_A^2\omega^*$
(see Equation \Eq{suz} above), we get 
\beq{sd_eq}
I\dot \omega^*=Z \dot M R_A^2(\omega_m-\omega^*)+z \dot M R_A^2\omega^*\,.
\eeq
It is obvious that for angular momentum removal from the neutron star through a shell , the coefficient $Z$ has to be larger than $z$. Then the accreting neutron star can episodically spin down (below we will explain this statement in more detail). And conversely, if $Z<z$, the neutron star can only spin up.

If a hot shell is not formed above the magnetosphere (at high X-ray luminosities or low velocity stellar winds, see e.g.  \cite{Sunyaev78} and references below), then the supersonic or Bondi accretion regime is established and no angular momentum can be removed from the neutron star. In this case  $Z=z$, equation \Eq{sd_eq} takes the simple form $I\dot \omega^*=Z \dot M R_A^2\omega_m $, and the neutron star will spin up to a frequency of the order of  $\omega_K(R_A)$ regardless of the sign of the difference between the angular momentum of the matter and the magnetic field lines $\omega_m-\omega^*$ close to the magnetospheric boundary. Due to conservation of the specific angular momentum $\omega_m=\omega_B (R_B/R_A)^2$. Without the presence of a shell the evolution of the angular frequency of the neutron star can be described by the equation 
\beq{}
I\dot \omega^*=Z \dot M \omega_BR_B^2\,,
\eeq 
where the coefficient $Z$ plays the role of the specific angular momentum of the matter. For example, in the models of \cite{IllarionovSunyaev75} $Z\simeq 1/4$. Numerical modeling of Bondi-Hoyle-Littleton accretion in two-dimensional (e.g.  \cite{FryxellTaam88, Ho_ea89}) and three-dimensional (e.g. \cite{Ruffert97, Ruffert99}) calculations have, however, shown that due to inhomogeneities in the stellar wind, accretion becomes non-stationary and the sign of the captured angular momentum may change. The sign of  $Z$  may thus also be negative and we may observe alternating spin-up and spin-down episodes. Such a scenario is often used to explain the observed changes in the sign of the torque in accreting X-ray pulsars (see the discussion in \cite{Nelson_ea97}). We stress again that this picture is completely realistic for X-ray pulsars at high luminosities  $>4\times 10^{36}$~erg/s, when due to the strong Compton cooling around the rotating magnetosphere no convective quasi-hydrostatic shell can be formed.

If a hot shell is indeed formed (at moderate X-ray luminosities less than $\sim 4\times 10^{36}$~erg/s, see \Eq{M*}), the angular momentum from the neutron star can be transferred outside through the convective shell by means of turbulent viscosity. Therefore, substituting $\omega_m$ from \Eq{omega_m1} and \Eq{sd_eq}, we get
\beq{sd_eq1}
I\dot \omega^*= Z\dot M \tilde\omega\omega_B R_B^2\myfrac{R_A}{R_B}^{2-n}-Z(1-z/Z)\dot M R_A^2\omega^*\,.
\eeq
This is the main formula that we will use in the following to describe the evolution of the spin of the neutron star.

The dimensionless coefficients in this equation can be calculated using the factor $f(u)$, which is included in the expressions for $Z$ and $R_A$. Thus, the only dimensionless parameter in the model is  $\myfrac{K_1}{\zeta}$. Below we will show how this coefficient can be determined using observational data from real X-ray pulsars.

\section{Spin-up and spin-down of X-ray pulsars}
\label{s:main}  

In this section we will study the dependence of the accelerating and decelerating torques on the accretion rate $\dot M$. We would like to stress, again, that in our case accretion is subsonic and the accretion rate is determined by the ability of matter to enter the magnetosphere through the shell. The velocity with which the plasma enters the magnetosphere is then mainly dependent on the density at the magnetospheric boundary. The density distribution in the shell is on the other hand directly connected to the density of matter in the shockwave region and density variations downstream the shock are thus rapidly translated to corresponding variations in the density near the magnetospheric boundary. This means that variations of the accretion rate onto neutron stars in binary systems with circular or low-eccentricity orbits should be essentially independent of orbital phase, and be mostly determined by variations in the stellar wind. In constrast, possible changes in the capture radius $R_B$ (for example due to velocity changes in the stellar wind or variations in the orbital velocity of the neutron star) have little effect on the accretion rate through the shell, but strongly affect the torques applied to the neutron star (see Equation \Eq{sd_eq1}). 

Equation \Eq{sd_eq1} can be rewritten in the form explicitely showing spin-up and spin-down torques:
\beq{sd_eq2}
I\dot \omega^*=A\dot M^{\frac{2n+3}{11}} - B\dot M^{3/11}\,.
\eeq

For a characteristic value of the accretion rate $\dot M_{16}\equiv \dot M/10^{16}$~g/s, the coefficients (not dependent on the accretion rate) will be equal to (in CGS units):
\beq{A(Z)}
A\approx 4.22\times 10^{31} (0.0388)^{2-n}\tilde \omega \myfrac{K_1}{\zeta}K_0^{-\frac{2n+3}{11}}
K_t^{\frac{9+6n}{22}}
\mu_{30}^{\frac{13-6n}{11}}\myfrac{v_8}{\sqrt{\delta}}^{-2n}\myfrac{P_b}{10\hbox{d}}^{-1}
\eeq
\beq{B}
B\approx 5.47\times 10^{32}(1-z/Z)\ \myfrac{K_1}{\zeta}K_0^{-3/11}K_t^{{9}/{22}}\mu_{30}^{{13}/{11}}\myfrac{P^*}{100\hbox{s}}^{-1}
\eeq
(From now on we will assume $\gamma=5/3$ in all numerical estimates.) The dimensionless factor $\delta<1$ takes into account the actual location of the gravitational capture radius, which for a cold stellar wind may be somewhat smaller than the Bondi radius \cite{Hunt71}. The capture radius can also be reduced due to radiative heating of the stellar wind by the X-rays from the neutron star (see below). To derive numerical values of the coefficients in Equations \Eq{A(Z)} and \Eq{B}, we used the expressions for the coefficient $Z$ \Eq{Zdef} using \Eq{fu} and \Eq{RA} for the Alfv\'en radius.

Below we will study the case $Z-z>0$, i.e. $B>0$, since in the opposite case only spin-up of the neutron star is possible.

\subsection{Equilibrium pulsars}
\label{s:eq}
 
For equilibrium pulsars we set 
$\dot \omega^*=0$ and from Equation \Eq{sd_eq} we get
\beq{Zeq}
Z_{eq}(\omega_m-\omega^*)+z\omega^*=0\,.
\eeq 

Close to equilibrium we may vary \Eq{sd_eq} with respect to $\dot M$. It is convenient to introduce the dimensionless parameter $y\equiv \dot M/\dot M_{eq}$, so that close to equilibrium $y=1$. Variations in $\delta \dot M$ may in general be caused by changes in density $\delta \rho$ as well as in velocity of the stellar wind $\delta v$ (and thus the Bondi radius). From the continuity equation and taking into account the dependence of $f(u)$ on $\dot M$ in the shell \Eq{fu}, 
we get
\beq{varrhov}
\frac{7}{11}\frac{\delta \dot M}{\dot M}=\frac{\delta \rho}{\rho}-3\frac{\delta  v}{v}
\eeq 
Let us start by studying variations in the density only. Assuming $R_B=const$, we find
\beq{deriv}
I\frac{\partial \dot\omega^*}{\partial \dot M}|_{eq}=
I\frac{1}{\dot M_{eq}}\frac{\partial \dot\omega^*}{\partial y}|_{y=1}=
\frac{4}{11}z\omega^*R_A^2+\frac{2n}{11}Z_{eq}\omega_mR_A^2
\eeq
Using the expression for $\omega_m$ from \Eq{Zeq} and substituting it into \Eq{deriv}, we get
\beq{Zeqrho}
Z_{eq,\rho}-\frac{n-2}{n}z=\frac{I\frac{\partial \dot\omega^*}{\partial \dot M}|_{eq}}{\frac{2n}{11}\omega^*R_A^2}\approx \frac{3.64}{n}\myfrac{\frac{\partial \dot\omega^*}{\partial y}|_{y=1}}{10^{-12}}\myfrac{P^*}{100s}K_0^{-4/11}K_T^{6/11}\dot M_{16}^{-7/11}\mu_{30}^{-12/11}\,.
\eeq
Now let us keep the density constant and study changes in the velocity only. Then, we have from \Eq{varrhov} that $\delta  v/v=-(7/33)\delta \dot M/\dot M$. Varying \Eq{sd_eq}, we get
\beq{Zeqv}
Z_{eq,v}-\frac{5n-3}{5n}z=\frac{I\frac{\partial \dot\omega^*}{\partial \dot M}|_{eq}}{\frac{20n}{33}\omega^*R_A^2}\approx \frac{1.1}{n}\myfrac{\frac{\partial \dot\omega^*}{\partial y}|_{y=1}}{10^{-12}}\myfrac{P^*}{100s}K_0^{-4/11}K_T^{6/11}\dot M_{16}^{-7/11}\mu_{30}^{-12/11}\,.
\eeq 

A majority of neutron stars in X-ray pulsars rotate close to their equilibrium periods, i.e. on 
average $\dot\omega^*=0$. Near equilibrium we get from \Eq{sd_eq2} in the settling accretion regime:
\beq{mu_eq}
\mu_{30}^{(eq)}\approx
\myfrac{0.077\cdot (0.0388)^{(2-n)}\tilde\omega}{1-z/Z}^\frac{11}{6n} K_0^{-1/3}K_t^{1/2}
\myfrac{\sqrt{\delta}}{v_8}^\frac{11}{3} \dot M_{16}^{1/3}
\myfrac{P_*/100 \hbox{s}}{P_b/10 \hbox{?}}^\frac{11}{6n}
\,.
\eeq
 
This expression can be reversed to give the equilibrium period for a system if the magnetic field is known:
\beq{Peq}
P_{eq}\approx \frac{1300 [\hbox{s}]}{0.0388^{2-n}}(1-z/Z_{eq})\tilde\omega^{-1}K_0^{2n/11}K_t^{-3n/11}\mu_{30,eq}^{6n/11}\myfrac{P_b}{10\hbox{d}}
\dot M_{16}^{-2n/11}\myfrac{v_8}{\sqrt{\delta}}^{2n}\,.
\eeq

The ratio of pulsar's to Keplerian frequency 
at the Alfv\'en radius is independent of $n$ and equal to 
\beq{e:omegatoomegaK}
\frac{\omega^*}{\omega_K(R_A)}\approx 0.27 K_0^{3/11}K_t^{-9/22}
\myfrac{P^*}{100\hbox{s}}^{-1}\mu_{30}^{9/11}\dot M_{16}^{-3/11}\,.
\eeq
At equilibrium, the ratio between the toroidal and polodial magnetic fields at the Alfv\'en radius (Equation \Eq{btbp}) takes the form: 
\beq{BtBpnum}
\frac{B_t}{B_p}|_{eq}=-\myfrac{K_1}{\zeta}\myfrac{z}{Z_{eq}}\myfrac{\omega^*}{\omega_K(R_A)}=
\frac{10f(u)z}{\sqrt{2} K_t}\myfrac{\omega^*}{\omega_K(R_A)}\,.
\eeq
Substituting $f(u)$ and \Eq{e:omegatoomegaK} in this expression, we get:
\beq{e:BtBp}
\left|\frac{B_t}{B_p}\right|_{eq}\approx 0.75z \frac{K_0^{10/11}}{K_t^{15/11}}
\myfrac{P^*}{100\hbox{s}}^{-1}\mu_{30}^{8/11}\dot M_{16}^{1/11}
\eeq

We stress that for slowly rotating accreting pulsars the ratio between the neutron star spin frequency and the Keplerian frequency at the Alfv\'en radius is always smaller than unity. Therefore, for typical values $f(u)\sim 0.3$ and $z=2/3$ we have $B_t/B_p<1.5 (\omega^*/\omega_K(R_A))<1$, and the pulsars are far from being in the propeller regime (see further discussion in Section \ref{s:propeller}).  

We would like to stress that in the important case $n=2$ (iso-angular-momentum distribution), the coefficient in the second term in \Eq{Zeqrho} vanishes, and thus equating $Z_{eq}$ to\Eq{Znum} we find the value of the magnetic moment of the neutron star only from the pulsar equilibrium period and the derivative $(\partial \dot\omega/\partial y)_{eq}$:
\beq{mueqnew}
\mu_{30,eq}\approx 5 \myfrac{\frac{\partial \dot\omega^*}{\partial y}|_{y=1}}{10^{-12}}\myfrac{P^*}{100s}\myfrac{K_1}{\zeta}^{-1}K_0^{3/11}K_t^{-3/7}\dot M_{16}^{-3/11}\,.
\eeq 
 
For the case $n=2$ and a known $\mu_{eq}$ we obtain the stellar wind velocity:
\beq{e:v8min}
\frac{v_8}{\sqrt{\delta}}\approx 0.53 (1-z/Z_{eq})^{-1/4}
K_0^{-1/11}K_t^{3/22} \dot M_{16}^{1/11}\mu_{30,eq}^{-3/11}
\myfrac{P_*/100 \hbox{s}}{P_b/10 \hbox{d}}^{1/4}\,.
\eeq

As will be shown below, for real equilibrium pulsars $z/Z_{eq}\ll 1$, and thus the derived formula gives a correct estimate of the stellar wind velocity. Note the weak dependence of the formula on the dimensionless constant as well as on the accretion rate. In the framework of our model we may thus, with knowledge of the equilibrium spin period $P^*$, the binary period $P_b$ and with an estimate of the neutron star magnetic field $\mu$, determine the stellar wind velocity, without complicated spectroscopic measurements.

\subsection{Non-equilibrium pulsars}
\label{s:noneq}
 
 \begin{figure*}
\includegraphics[width=0.5\textwidth]{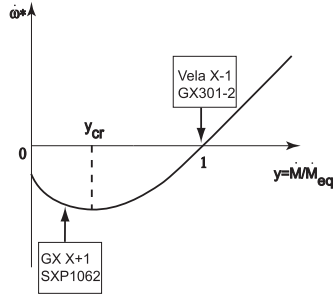}
\caption{An illustration of the dependence of $\dot\omega^*$ on the dimensionless accretion rate $y$ [\Eq{sdy}]. In fact as $y\to 0$, $\dot \omega^*$ approaches some negative value since the neutron star 
enters the propeller regime at small accretion rates. The figure shows the position in the diagram for equilibrium pulsars with $y\sim 1$ and for non-equilibrium pulsars at steady spin-down with $y<y_{cr}$}
\label{f:y}
 \end{figure*}
 
Below we will study the case $Z-z>0$, and thus $B>0$, since in the opposite case only spin-up is possible.

First of all, we note that the function $\dot\omega^*(\dot M)$ reaches a minimum for some $\dot M_{cr}$. Differentiating equation \Eq{sd_eq2} with respect to $\dot M$ and equating the achieved expression to zero, we find
\beq{dotMcr}
\dot M_{cr}=\left[\frac{B}{A}\frac{3}{(3+2n)}\right]^{\frac{11}{2n}}\,.
\eeq 
For $\dot M=\dot M_{cr}$ the expression $\dot\omega^*$ reaches an absolut minimum (see. Fig.\ref{f:y}).

It is convenient to introduce the dimensionless parameter
\beq{y_def}
y\equiv \frac{\dot M}{\dot M_{eq}}
\eeq
where $\dot M_{eq}$ represents the accretion rate at which $\dot\omega^*=0$:
\beq{dotmeq}
\dot M_{eq}=\myfrac{B}{A}^{11/2n}
\eeq
Obviously,
\beq{}
\dot M_{cr}=\dot M_{eq} \myfrac{3}{2n+3}^\frac{11}{2n} \,,
\eeq
In other words, $\dot\omega^*$ has a minimum for a value of the dimensionless parameter of
\beq{ycr}
y_{cr}=\myfrac{3}{2n+3}^\frac{11}{2n}<1.
\eeq
Equation \Eq{sd_eq2} can be rewritten in the form 
\beq{sdy}
I\dot \omega^*=A\dot M_{eq}^\frac{3+2n}{11}y^\frac{3+2n}{11}
\left(1- y^{-\frac{2n}{11}}\right)\,,
\eeq

The minimum $\dot \omega^*$ for $y=y_{cr}$ (i.e. the maximum possible spin-down rate of the pulsar) is
\beq{omegadotmin}
I\dot \omega^*_{min}=-\frac{2n}{3}A\dot M_{eq}^\frac{3+2n}{11}y^\frac{3+2n}{11}\,.
\eeq 

Now, we vary \Eq{sdy} with respect to $y$:
\beq{variations}
I(\delta\dot\omega^*)=I\frac{\partial \dot \omega^*}{\partial y}(\delta y)=
\frac{3}{11}A\dot M_{eq}^\frac{3+2n}{11}y^{-8/11}
\left(\frac{2n+3}{3}y^\frac{2n}{11}-1\right)(\delta y)\,.
\eeq
Apparently, depending on whether $y>y_{cr}$ or $y<y_{cr}$, 
\textit{correlated changes} of $\delta \dot\omega^*$
with X-ray flux should have different signs. Indeed, for GX 1+4 in \cite{Chakrabarty_ea97} and
\cite{GonzalezGalan_ea12} a positive correlation
of the observed $\delta P$ with $\delta \dot M$ was found using the CGRO \textit{BATSE}
and \textit{Fermi} GBM data. This means that 
there is a negative correlation between $\delta\omega^*$ and $\delta\dot M$, suggesting $y<y_{cr}$ in this source.

Let us now consider accreting pulsars in the stage of spin-down (like e.g. GX 1+4 and SXP 1062). If the pulsar is spinning down, measurements of the spin-down rate give limits on the parameters of our model. From the simple fact that the spin down is stable, using equations \Eq{sd_eq2}, \Eq{A(Z)} and \Eq{B} we may obtain a lower limit on the magnetic field in the case of quasi-spherical accretion with $\dot\omega^*<0$,
\beq{e:mulim}
\mu_{30}>\mu_{30, min}\approx 0.1
(1-z/Z)^{-\frac{11}{12}} \tilde\omega^\frac{11}{12}K_0^{-1/3}K_t^{1/2}
\myfrac{\sqrt{\delta}}{v_8}^\frac{11}{3} \dot M_{16}^{1/3}
\myfrac{P_*/100 \hbox{s}}{P_b/10 \hbox{d}}^\frac{11}{12}
\eeq 
(and thus equation \Eq{mu_eq} is here transformed into an inequality). We now make use of the fact that during spin down there is a maximum possible breaking torque (see equation \Eq{omegadotmin}). Inserting the values of the coefficients $A$ and $B$ from equations \Eq{A(Z)} and \Eq{B} into \Eq{omegadotmin}, we find:
\beq{e:omegadotsdmax}
\dot\omega^*_{sd,max}\approx -1.13\times 10^{-12}[\hbox{rad/s}] (1-z/Z)^{7/4} \myfrac{K_1}{\zeta} 
\mu_{30}^{2}\myfrac{v_8}{\sqrt{\delta}}^{3}\myfrac{P^*}{100\hbox{s}}^{-7/4}
\myfrac{P_b}{10\hbox{d}}^{3/4}\,.
\eeq
For the accretion rate $\dot M=\dot M_{cr}$ this expression reaches the numerical value 
\beq{dotmcr}
\dot M_{16,cr}\approx 112 (1-z/Z)^{11/4}K_0K_t^{-2}\mu_{30}^3\myfrac{v_8}{\sqrt{\delta}}^{11}
\myfrac{P_b/10 \hbox{d}}{P_*/100 \hbox{s}}^\frac{11}{4}.
\eeq
(Note the extremely strong dependence on the stellar wind velocity.) 

Then, from the condition $|\dot \omega^*_{sd}|\le |\dot\omega^*_{sd,max}|$ follows a more interesting lower limit on the neutron star magnetic field:
\beq{e:mulim1}
\mu_{30}>\mu_{30, min}'\approx 0.94
\left|\frac{\dot\omega^*_{sd}}{10^{-12}\hbox{rad/s}}\right| \myfrac{K_1}{\zeta}^{-1/2}
\myfrac{v_8}{\sqrt{\delta}}^{-3/2}\myfrac{P^*}{100\hbox{s}}^{7/8}
\myfrac{P_b}{10\hbox{d}}^{-3/8}.
\eeq
Note the weaker dependence of this estimate on the stellar wind velocity as compared to the inequality \Eq{e:mulim}.

If the accelerating torque can be neglected compared to the breaking torque (corresponding to the low X-ray luminosity limit $y\ll 1$), we find directly from \Eq{sd1} that for accreting pulsars at spin down,  
\beq{e:sdonly}
\dot \omega^*_{sd}\approx - 0.55\times 10^{-12}[\hbox{rad/s}]\myfrac{K_1}{\zeta}K_0^{-3/11}K_t^{9/22} \mu_{30}^{13/11}\dot M_{16}^{3/11}\myfrac{P^*}{100\hbox{s}}^{-1}.
\eeq
From this we obtain a lower limit on the neutron star magnetic field that does not depend on the parameters of the stellar wind nor the binary orbital period:
\beq{e:mulim2}
\mu_{30}>\mu_{30, min}''\approx 1.66 
\left|\frac{\dot\omega^*_{sd}}{10^{-12}\hbox{rad/s}}\right|^{11/13} \myfrac{K_1}{\zeta}^{-11/13}K_0^{3/13}K_t^{-9/26}\dot M_{16}^{-3/13}
\myfrac{P^*}{100\hbox{s}}^{11/13}\,.
\eeq 
 
Eliminating $\myfrac{K_1}{\zeta}$ from  \Eq{btbpK1} and \Eq{sd1} we get:
\beq{e:BtBplim}
\left|\frac{B_t}{B_p}\right|=\tilde K\left|\frac{I\dot\omega^*_{sd}R_A^3}{ K_2\mu^2}\right|
\approx 0.49 \left|\frac{\dot\omega^*_{sd}}{10^{-12}\hbox{rad/s}}\right|\mu_{30}^{-4/11}K_0^{6/11}K_t^{-9/11}\dot M_{16}^{-6/11}\,.
\eeq
We see from \Eq{e:BtBplim}, that with decreasing $\dot M$ the ratio $B_t/B_p$ increases for reasons well understood --- at low $\dot M$ the characteristic cooling time for the plasma increases and the toroidal component has time to grow to the same strength as the poloidal. $B_t$ can, however, not become larger than $B_p$ due to an instability similar to that of a tightly wound spring. Equating $B_t=B_p$, and using \Eq{e:BtBplim}, we find the luminosity below which the pulsar enters the strong coupling regime during spin down (see Section \ref{s:strongcoupling} above): 
\beq{mdotcr}
\dot M^*_{16}\approx 0.27  \left|\frac{\dot\omega^*_{sd}}{10^{-12}\hbox{rad/s}}\right|^{11/6}\mu_{30}^{-2/3}K_0K_t^{-3/2}\,.
\eeq
Below this luminosity in the strong coupling regime the spin-down law becomes $K_{sd}\sim \mu^2 R_A^{-3}\sim \dot M^{6/11}$:
\beq{Ksd611}
\dot \omega^*_{sd}\approx -2\times 10^{-12}[\hbox{rad/s}]
\mu_{30}^{4/11}K_0^{-6/11}K_t^{9/11}\dot M_{16}^{6/11}
\eeq
(Note that when the spin-up torque can be neglected the expression does not contain the - ever so hard to determine - velocity of the stellar wind. )

For a further decrease of the accretion rate in non-equilibrium pulsars, the Alfv\'en radius will grow to the corotation radius and the pulsar may enter a transient state (the propeller regime). From the condition $\omega^*=\sqrt{GM/R_A^3}$ we find the accretion rate for this transition:
\beq{dotm**}
\dot M^{**}_{16}\approx 0.0082 K_0 K_t^{-3/2}\mu_{30}^{3}
\myfrac{P^*}{100\hbox{s}}^{-11/3}\,.
\eeq

The formulae derived above show that the restrictions on the model become more significant if the neutron star magnetic field can be measured independently (for example using spectral cyclotron lines). We also would like to stress the fact that measurements of correlated fluctuations of the spin frequency derivative with luminosity during spin down allows us to place the source in a $\dot \omega^*-y$ diagram (see Fig. \ref{f:y}). To the right from the minimum $y>y_{cr}$ and the correlation positive. To the left $y<y_{cr}$ and the correlation is negative. This way we may obtain further limits on the parameters of our model. Below we will perform this analysis for the source GX 1+4, in which such correlations where measured \cite{Chakrabarty_ea97}, \cite{GonzalezGalan_ea12}.

\section{Application to real X-ray pulsars}
\label{s:applications} 

In this Section, as an illustration of the possible applicability of our model
to real sources, 
we will consider five particular slowly rotating moderately luminous X-ray pulsars: GX 301-2,
Vela X-1, GX 1+4, SXP 1062 and 4U 2204+56. The first two pulsars are close to the equilibrium rotation of the neutron star, showing spin-up/spin-down excursions near the equilibrium frequency
(apart from the spin-up/spin-down jumps, which may be, we think, due to episodic 
switch-ons of the strong coupling regime when the toroidal magnetic field component
becomes comparable to the poloidal one, see Section \ref{s:strongcoupling}). The third source, GX 1+4, is a typical example of a pulsar displaying long-term spin-up/spin-down episodes.
During the last 30 years, it has shown a steady spin-down with frequency fluctuations (anti-)correlated with luminosity  (see \cite{GonzalezGalan_ea12} for a more detailed discussion). Clearly, this pulsar can not be considered to be in equilibrium. 
The pulsar SXP 1062 in the Large Magellanic Cloud as well as the pulsar 4U 2206+54 have only been observed at steady spin-down.

\subsection{GX 301-2}
GX301--2 (also known as 4U 1223--62) is a high-mass X-ray binary, consisting of a neutron star and an early type B optical companion with mass $\simeq 40 M_\odot$ and radius $\simeq 60 R_\odot$. The binary period is 41.5 days \cite{Koh_ea97}. 
The neutron star is a $\sim680$~s X-ray pulsar \cite{White_ea76}, accreting from the strong wind of its companion ($\dot M_{loss} \sim 10^{-5} \ms$/yr, \cite{Kaper_ea06}). The 
photospheric escape velocity of the wind is $v_{esc}\approx 500$~km/s. The semi-major
axis of the binary system is $a\approx 170 R_\odot$ and the orbital eccentricity $e\approx 0.46$. 
The wind terminal velocity was found  \cite{
Kaper_ea06} to be about 300 km/s, smaller than the photospheric escape velocity.  
 
GX 301-2 shows strong short-term pulse period variability, which, as in many other wind-accreting pulsars, can be well described by a random walk model  \cite{deKoolAnzer93}. Earlier observations between 1975 and 1984 showed a period of $\sim 700$~s while in 1984 the source started to spin up \cite{Nagase89}. The almost 10 years of spin-up were followed by a reversal of spin in 1993 \cite{PravdoGhosh01} after which the source has been continuously spinning down \cite{LaBarbera_ea05}, \cite{Kreykenbohm_ea04}, \cite{Doroshenko_ea10}. Rapid spin-up episodes sometimes appear in the \textit{Fermi} GBM data on top of the long-term spin-down trend \cite{Finger_ea11}.
It can not be excluded that these rapid spin-up episodes, as well as similar ones observed in BATSE data, reflect a temporary entrance into the strong coupling regime, as discussed in Section 2.4.1. Cyclotron line measurements \cite{LaBarbera_ea05} yield a magnetic 
field estimate near the neutron star surface of
$B_0\approx 4.4\times 10^{12}$~G ($\mu=1/2 B_0 R_0^3=2.2\times 10^{30}$~G cm$^3$ for
the assumed neutron star radius $R_0=10$~km).
 
\begin{figure*}
\includegraphics{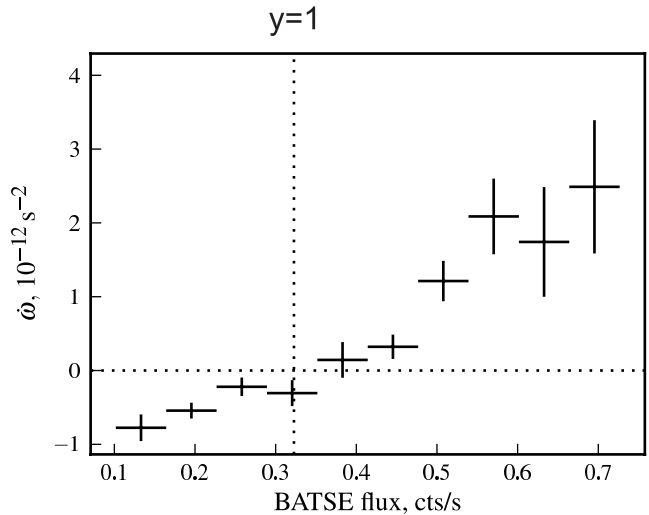}
\caption{Torque-luminosity correlation in GX 301-2, $\dot\omega^*$ as a function of 
BATSE data (20-40 keV pulsed flux) near the equilibrium frequency, see \cite{Doroshenko_ea10}. The assumed X-ray flux at equilibrium (in terms of
the dimensionless parameter $y$) is also shown by the vertical dotted line.}
\label{f:gx301}
 \end{figure*}
 
In Fig. \ref{f:gx301} we have plotted $\dot\omega^*$ as a function of the observed pulsed flux (20-40~keV) according to BATSE data (see \cite{Doroshenko_ea10} for more detail). We will consider the neutron star magnetic field in this source to be known from observations.
An estimate of $\dot M$ can be inferred from the X-ray flux provided 
the distance to the source is known, which is generally not the case to a great certainty.
We shall assume that near equilibrium a hot quasi-spherical shell exists
in this pulsar and that the accretion rate is $3\times 10^{16}$~g/s, i.e.
not higher than the critical value $\dot M_*\simeq 4\times 10^{16}$~g/s [\Eq{M*}].
The derivative $\partial \dot \omega^*/\partial y$
can be derived from the $\dot \omega^*$ -- X-ray flux plot, since 
in the first approximation the accretion rate is proportional to the observed pulsed X-ray flux. Near the equilibrium (the torque reversal point with $\dot \omega^*=0$), we find 
from a linear fit in Fig. \ref{f:gx301}   
$\partial \dot \omega^*/\partial y \approx 1.5 \times 10^{-12}$~rad/s$^2$. 

The obtained parameters ($Z$, $\myfrac{K_1}{\zeta}$ etc.) for this pulsar are listed in Table 1. We note that the toroidal component of the magnetic field is much less than the poloidal (the pulsar is far from the strong-coupling limit). The stellar wind velocity, determined using the formula \Eq{e:v8min}, is close to the photospheric escape velocity. We also note that the value of the parameter describing the coupling between the plasma and the magnetosphere $K_1/\zeta$ is of the order of 14, although by its physical sense the coefficient $K_1$ should be of the order of 1. This means that the value of the parameter $\zeta$, which gives the characteristic relative size of the region in which transfer of angular momentum from the shell takes place to the magnetosphere (or vice versa) has to be of the order of 1/10 (i.e. the characteristic size of the region where angular momentum transfer takes place should be approximately 1/10 of the Alfv\'en radius).

\subsection{Vela X-1}
 
\begin{figure*}
\includegraphics{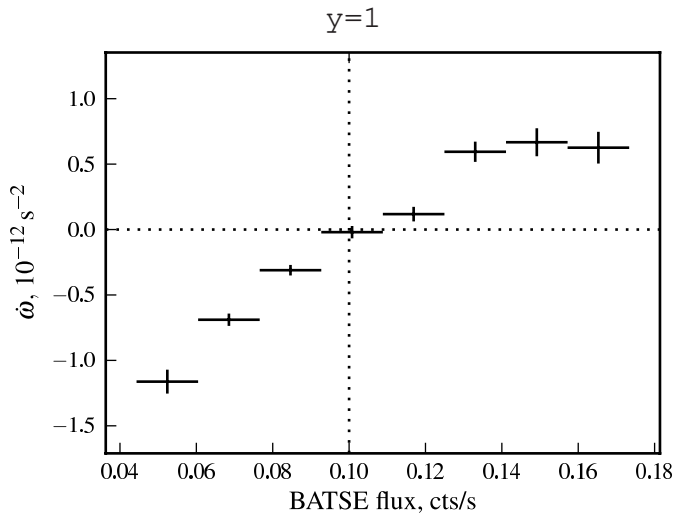}
\caption{The same as in Fig. \ref{f:gx301} for Vela X-1 \cite{Doroshenko11}.}
\label{f:velaX1}
 \end{figure*}

Vela X-1 (=4U 0900-40) is the brightest persistent accretion-powered pulsar in the 20-50 keV energy band with an average luminosity of $L_{x} \approx 4\times10^{36}$erg/s \cite{Nagase89}.
It consists of a massive neutron star (1.88 $\msun$, \cite{Quaintrell_ea03}) and the B0.5Ib super giant HD 77581, which eclipses the neutron star every orbital cycle of $\sim 8.964$ days \cite{vanKerkwijk_ea95}. The neutron star was discovered as an X-ray pulsar with a spin period of $\sim$283~s \cite{Rappaport75}, which has remained almost constant since the discovery of the source. The optical companion has a mass and radius of $\sim 23$ $M_\odot$ and $\sim 30$ $R_{sun}$ respectively  \cite{vanKerkwijk_ea95}. The photospheric escape velocity is $v_{esc}\approx 540$~km/s. The orbital separation is $a\approx 50 R_\odot$ and the orbital eccentricity $e\approx 0.1$. The primary almost fills its Roche lobe (as also evidenced by 
the presence of elliptical variations in the optical light curve,  \cite{Bochkarev_ea75}. The mass-loss rate from the primary star is $10^{-6}$ $\msun$/yr (Nagase et al. 1986) via a fast wind with a terminal velocity of $\sim 1100$~km/s \cite{Watanabe_ea06}, which is typical for this class. 
Despite the fact that the terminal velocity of the wind is rather large, the compactness of the system makes it impossible for the wind to reach this velocity before interacting with the neutron star, so the relative velocity of the wind with respect to the neutron star is rather low, $\sim 700$~km/s.

Cyclotron line measurements \cite{Staubert_03} yield the magnetic field estimate 
$B_0\approx 3\times 10^{12}$~G ($\mu=1.5\times 10^{30}$~G cm$^3$
for the assumed neutron star radius 10~km). We shall assume that in this pulsar  
$\dot M\simeq 3\times 10^{16}$~g/s (again for the existence of a shell to be possible).
In Fig. \ref{f:velaX1} we have plotted $\dot\omega^*$ as a function of 
the observed pulsed flux (20-40~keV) according to BATSE data 
 \cite{Doroshenko11}. As in the case of GX 301-2, from a linear fit we find at the spin-up/spin-down transition point 
$\partial \dot \omega^*/\partial y \approx 1.2\times 10^{-12}$~rad/s$^2$.  

The obtained parameters for Vela X-1 are listed in Table 1. We note that the velocity of the stellar wind as obtained using \Eq{e:v8min} is very close to the observed value of 700 km/s. 
As in the case of GX 1+4, the value of the coupling parameter $K_1/\zeta$ is of the order of 10, i.e. the size of the region for transfer of angular momentum between the plasma and the magnetosphere is about 1/10 of the Alfv\'en radius.

\subsection{GX 1+4}

GX 1+4 was the first source to be identified as a symbiotic binary containing a neutron star \cite{Davidsen_ea77}. The pulse period is $\sim 140$~s  and the donor is an MIII giant \cite{Davidsen_ea77}. The system has an orbital period of 1161 days \cite{Hinkle_ea06}, making it the widest known LMXB by at least one order of magnitude. The donor is far 
from filling its Roche lobe and accretion onto the neutron star is by capture of the stellar wind of the companion. 
 
The system has a very interesting spin history. During the 1970's it was spinning up at the fastest rate ($\dot \omega_{su} \sim 3.8\cdot 10^{-11}$ rad/s) among the known X-ray pulsars at the time (e.g. \cite{Nagase89})). After several years of non-detections in the early 1980's, it reappeared again, now spinning down at a rate similar in magnitude to that of the previous spin-up. At present the source is steadily spinning down with an average spin down rate of $\dot omega^*_{sd}\approx -2.34\times 10^{-11}$~rad/s.  The observed spin-reversal has been interpreted in terms of a retrograde accretion disc forming in the system \cite{Makishima_ea88}, \cite{Dotani_ea89}, \cite{Chakrabarty_ea97}. A detailed spin-down history of the source is discussed in the recent paper \cite{GonzalezGalan_ea12}. Using our model this behavior can, however, be readily explained in the framework of quasi-spherical accretion.

As the pulsar in GX 1+4 is not in equilibrium, we use one of the three formulas from Section \ref{s:noneq} to derive a lower limit on the neutron star magnetic field from the observed value of $\dot \omega_{sd}$. From \Eq{e:mulim1} we get $\mu_{30,min}'\approx 12 (K_1/\zeta)^{-1/2}$. Assuming that the coupling parameter for non-equilibrium pulsars is similar to that in equilibrium ones (and thus that the size of the region where transfer of angular momentum between the plasma and the magnetosphere takes place is of the order of 1/10 of the Alfv\'en radius, $\zeta\sim 0.1$) we find that $\mu_{30,min}'\sim 4$. 

In this source we also observe anti-correlated variability in spin-down rate versus X-ray luminosity \cite{Chakrabarty_ea97}. According to the latest \textit{Fermi} GBM data in the paper \cite{GonzalezGalan_ea12} it was found that $-\dot\omega^*\sim L_x^{0.3}$. In our model for moderate coupling, $K_{sd}\sim \dot M^{3/11}$, which is very similar to the observed relation. In the earlier \textit{BATSE} observations \cite{Chakrabarty_ea97} it was found that $-\dot\omega^*\sim L_x^{0.48}$. It can not be excluded that the average luminosity of the source was lower at this time. In that case the component  $B_t$ could have been closer to $B_p$, and then the expected correlation would have had the form $K_{sd}\sim \dot M^{6/11}\sim L_x^{0.54}$.
Note that in the model with weak coupling (with transfer of angular momentum due to turbulence close to the magnetosphere \cite{Lipunov87}), the breaking torque is less effective by a factor of $(R_A/R_c)^{3/2}$ and not at all dependent on the luminosity. In low-luminosity pulsars the cooling close to the Alfv\'en radius is less effective, which leads to the development of convective movements in the shell and the establishment of the moderate coupling regime.

Further, we note that the short-term spin-up episodes, sometimes observed on top
of the steady spin-down behaviour (at about MJD 49700, see Fig. 2 in \cite{Chakrabarty_ea97} ) are 
correlated with an enhancement of the X-ray flux, in contrast to the negative frequency-flux correlations 
discussed above. During these short spin-ups, $\dot\omega^*$ is about half the average 
$\dot\omega^*_{su}$ observed during the steady spin-up state of GX 1+4 up to 1980. The X-ray luminosity during these episodic spin-ups is approximately five times 
larger than the mean X-ray luminosity during the steady spin-down. We remind the reader that 
once $\dot M>\dot M_*$, a free-fall gap appears above the magnetosphere, and the
neutron star can only spin up. When the X-ray flux drops again, the settling accretion 
regime is re-established and the neutron star resumes its spinning-down.

\begin{table*}
\label{T2}
 \centering
 \caption{Parameters for the pulsars discussed in Section \ref{s:applications}. References for the observed pulsar and orbital parameters are given in the text as well as values for the wind velocities from measurements of the optical components. The parameters $Z$, $K_1/\zeta$ and $f(u)$ were derived in Sections \ref{s:f(u)} and \ref{s:angmom}.
Numerical estimates are given assuming iso-angular-momentum rotation in the shell ($n=2$), moderate coupling between the plasma and the magnetic field $\delta=1, \zeta=1$, $\tilde\omega=1$, $K_0=1$, $\gamma=5/3$ without turbulence ($m_t=0$, $K_t=1$).} 
 $$
\begin{array}{lcc|ccc}
\hline
\hbox{Pulsars }&\multicolumn{2}{c}{\hbox{equilibrium pulsars }} &
\multicolumn{3}{c}{\hbox{non-equilibrium pulsars }}\\
\hline
& {\rm GX 301-2} & {\rm Vela X-1} 
& {\rm GX 1+4} &{\rm SXP1062}&{\rm 4U 2206+54}\\
\hline
\multicolumn{5}{c}{\hbox{Measured parameters}}\\
\hline
P^*{\hbox{(s)}} & 680 & 283 & 140 & 1062 &5560\\
P_B {\hbox{(d)}} & 41.5 & 8.96 & 1161 & \sim 300^\dag& 19\\
v_{w} {\hbox{(km/s)}} & 300 & 700 & 200 & \sim 300^\ddag& 350\\
\mu_{30}& 2.7 & 1.2 & ? & ? & 1.7\\
\dot M_{16} & 3 & 3 & 1 & 0.6 & 0.2\\ 
\frac{\partial \dot \omega}{\partial y} \arrowvert_{y=1}{\hbox{(rad/s}^2)} 
& 1.5\cdot10^{-12} & 1.2\cdot10^{-12} & n/a & n/a & n/a \\
 \dot\omega^*_{sd} & 0 & 0 & - 2.34 \cdot 10^{-11} & - 1.63 \cdot 10^{-11} & -9.4 \cdot 10^{-14}\\
\hline
\multicolumn{5}{c}{\hbox{Derived parameters}}\\
\hline
f(u) & 0.53 & 0.57 \\
\myfrac{K_1}{\zeta}& 14 &10& & & \gtrsim 8\\
Z& 3.7 & 2.6\\
B_t/B_p & 0.17 & 0.22\\
R_A{\hbox{(cm)}}& 2\cdot 10^9 & 1.4\cdot 10^9\\
\omega^*/\omega_K(R_A)& 0.07 & 0.08\\
v_{w,min} \hbox{(km/s)}& 500 & 740\\
\mu_{30,min}& & &\mu_{min}'\approx4&\mu_{min}''\approx20&\mu_{min}'\approx 3.6 \\
\hline
\end{array}
$$
$^\dag$ Estimate of the source's position in the Corbet diagram
$^\ddag$ Estimate of typical wind velocity binary pulsars containing Be-stars.
\end{table*}

\subsection{SXP 1062}
This recently discovered young X-ray pulsar in Be/X-ray binary system, located in a supernova remnant in the Small Magellanic Cloud. Its rotational period is $P^*\approx 1062$~s and it has a low X-ray luminosity of $L_x\approx 6\times 10^{35}$~erg/s \cite{HenaultBrunet_ea12}. The source shows a remarkably high spin-down rate of $\dot \omega^*\approx -  1.6\times 10^{-11}$~(rad/s$^2$). Its origin is widely discussed in the literature (see e.g. \cite{Haberl_ea12}, \cite{PopovTurolla12}) and a possibly anormously high magnetic field of the neutron star has been suggested \cite{FuLi12}. In the framework of our model we use more conservative limits. 
Neglecting the spin-up torque \Eq{e:mulim2}, we get $\mu_{30}>\mu_{30,min}''\approx 20$. This shows that the observed spin down can be explained by a magnetic field of the order of $10^{13}$~G, and thus we believe it is premature to conclude that the source is an accreting magnetar.

\subsection{4U 2204+56}

This slowly rotating pulsar has a period of $P^*=5560$~s and shows a spin-down rate of $\dot \omega_{sd}\approx -9.4 \times 10^{-14}$~rad/s \cite{Reig_ea12}. The orbital period of the binary system is $P_b\simeq 19$~days \cite{Reig_ea12}, and the measured stellar wind velocity is $v_W\approx 350$~km/s, abnormally low for an O9.5V \cite{Ribo_ea06} optical counterpart. The X-ray luminosity of the source is on average $L_x\simeq 2\times 10^{35}$~erg/s. A feature in the X-ray spectrum sometimes observed around 30 keV can be interpreted as a cyclotron line \cite{Torrejon_ea04}, \cite{Masetti_ea04}, \cite{Blay_ea06}, \cite{Wang09}. That gives an estimate of the magnetic field of the order of $B\sim (30/11.6)\times 1.3\approx  3.4\times 10^{12}$~G (taking into account the gravitational redshift close to the surface $1+z\sim 1.3$), and thus $\mu_{30}\approx 1.7$. Using this value of the magnetic field and neglecting the accelerating torque, from the formula in \Eq{e:sdonly} we obtain a lower limit on the parameter $K_1/\zeta\gtrsim 8$, which is very close to the coupling parameter values for the equilibrium pulsars Vela X-1 and GX 301-2.  
If we consider the magnetic field to be unknown (see discussion in \cite{Reig_ea12}), and apply the formula\Eq{e:mulim1}, like in the case of GX 1+4, assuming moderate coupling with $K_1/\zeta\sim 10$, we get the limit $\mu_{30}>\mu_{30,min}'\approx 3.6$, which is in agreement with standard neutron star magnetic field values. Note that  using our formulas for equilibrium pulsars would here give a magnetar value for the magnetic field \cite{Reig_ea12}.

\section{Discussion}
\label{s:discussion} 

\subsection{Physical conditions inside the shell}
For an accretion 
shell to be formed around the neutron star magnetosphere it is necessary that 
the matter crossing the bow shock does not cool down too rapidly and thus starts to fall freely. 
This means that the radiation cooling time $t_{cool}$ must be longer than the characteristic 
time of plasma motion. 

The plasma is heated up in the strong shock to a temperature
\beq{T_ps}
T_{ps}=\frac{3}{16}\mu_m\frac{ v_w^2}{\cal R}\approx 1.36\times 10^5 
[\hbox{K}]\myfrac{v_w}{100 \hbox{km/s}}^2\,.
\eeq
The radiative cooling time of the plasma is
\beq{t_cool}
t_{cool}=\frac{3kT}{2\mu_m n_e \Lambda}
\eeq 
where $\rho$ is the plasma density, $n_e=Y_e\rho/m_p$ is the electron number density 
( $\mu_m=0.6$ and $Y_e\approx 0.8$ for fully ionized plasma 
with solar abundance).
$\Lambda$ is the cooling function which can be approximated as 
\begin{equation}
\label{mcore} \Lambda (T)=\left\{
\begin{array}{l}
0, T<10^4 \, {\rm K} \\
1.0\times 10^{-24} T^{0.55} , 10^4 \, {\rm K} <T<10^5 \, {\rm K}   \\
6.2\times 10^{-19} T^{-0.6} , 10^5 \, {\rm K} <T<4\times 10^7 \, {\rm K}  \\
2.5\times 10^{-27} T^{0.5} , T>4\times 10^7 \, {\rm K}
\end{array}
\right. 
\end{equation}
\cite{Raymond_ea76}, \cite{Cowie_ea81}.  

Compton cooling becomes effective from the radius where the gas temperature $T$,
determined by the hydrostatic formula \Eq{hse_sol}, is lower than the X-ray Compton 
temperature $T_x$. The Compton cooling time (see \Eq{t_comp}) is:
\beq{t_C1}
t_{C}\approx 1060[\hbox{s}] \dot M_{16}^{-1}\myfrac{R}{10^{10}\hbox{cm}}^2\,.
\eeq
 
Above the
radius where $T_x=T$, Compton heating dominates. 
Taking the actual temperature close to the adiabatic one [\Eq{hse_sol}], 
we find $R_x\approx 2\times 10^{10}$~cm. 
We note that both the Compton 
and photoionization heating processes are controlled by the photoionization parameter
$\xi$ \cite{Tarter_ea69}, \cite{Hatchett_ea76}
\beq{ksi}
\xi=\frac{L_x}{n_eR^2}\,.
\eeq 

In most part of the accretion flux, $n\sim R^{-3/2}$, so $\xi\sim R^{-1/2}$ and
independent of the X-ray luminosity through the mass continuity equation. 
We derive a characteristic value for $\xi$: 
\beq{kxi_n}
\xi\approx 5\times 10^5 f(u) R_{10}^{-1/2}\,.
\eeq 
If Compton processes were effective everywhere, this high value of the parameter $\xi$
would imply that the plasma is Compton-heated up to keV-temperatures out to very large 
distances $\sim 10^{12}$~cm. However, at large distances the Compton heating time 
becomes longer than the characteristic time of gas accretion:
\beq{}
\frac{t_{C}}{t_{accr}}=\frac{t_{C}f(u)u_{ff}}{R}\approx 20 f(u) \dot M_{16}^{-1}R_{10}^{1/2}\,,
\eeq 
which shows that Compton heating is ineffective. The gas temperature is determined
by photoionization heating only and 
the gas can only be heated up to $T_{max}\approx 5\times 10^5$~K \cite{Tarter_ea69}, which is substantially lower than $T_x\sim 3$~keV.  

The effective gravitational capture radius corresponding to the sound velocity of the gas in the photoionization-heated zone is 
\beq{R_BC}  
R_{B}^*=\frac{2GM}{c_s^2}=\frac{2GM}{\gamma{\cal R} T_{max}/\mu_m}\approx 3.5\times 10^{12}\hbox{cm}
\myfrac{T_{max}}{5\times 10^5\hbox{K}}^{-1}\,.
\eeq
Everywhere up to the bow shock photoionization keeps the temperature at a value 
$\simeq T_{max}$. The sound velocity corresponding to $T_{max}$ is approximately 80 km/s. If the stellar wind velocity exceeds $80$~km/s, a standard bow shock is formed at the Bondi radius with a post-shock temperature given by \Eq{T_ps}. If the stellar wind velocity is lower than this value, the shock disappears and quasi-spherical accretion occurs from $R_B^*$.  
The photoionization heating time at the effective Bondi radius $3\times 10^{12}$~cm  is  
\beq{}
t_{pi}\approx \frac{(3/2)kT_{max}/\mu_m}{(h\nu_{eff}-\zeta_{eff})n_\gamma \sigma_{eff}c}
\approx 2\times 10^4 [\hbox{s}] \dot M_{16}^{-1}\,.
\eeq
(here $h\nu_{eff}\sim 10$~keV is the characteristic photon energy, $\zeta$ is the effective photoionization potential, $\sigma_{eff}\sim 10^{-24}$~cm$^2$ is the typical photoionization cross-section and $n_\gamma=L/(4\pi R^2 h\nu_{eff} c)$ is the photon number
density). The photoionization to accretion time ratio at the effective Bondi radius is then
\beq{}
\frac{t_{pi}}{t_{accr}}\approx 0.07 f(u) \dot M_{16}^{-1}\,.
\eeq 

At wind velocities $v_w>80$ km/s the bow shock stands at the classical Bondi radius $R_B$
inside the effective Bondi radius
$R_B^*$
determined by \Eq{R_BC}. The cooling time of the shocked
plasma at $R_B$ expressed through the wind velocity $v_w$ is:  
\beq{t_cool1}
t_{cool}\approx 4.7\times 10^4 [\hbox{s}]\dot M_{16}^{-1} v_7^{0.2}\,.
\eeq 
The photoionization heating time in the post-shock region can also be expressed through the
stellar wind velocity: 
\beq{}
t_{pi}\approx 3.5 \times 10^4 [\hbox{?}]\dot M_{16}^{-1} v_7^{-4}\,.
\eeq
A comparison of these two characteristic timescales implies that for low wind velocities radiative cooling becomes important and the source enters the regime of free-fall accretion with conservation of specific angular momentum.

Thus, for low wind velocities the plasma behind the shock cools down and starts to fall freely. As the cold plasma approaches
the gravitating center, photoionization heating becomes important and rapidly heats up the plasma to $T_{max}\approx 5\times 10^5$~K. Should this occur at a radius where $T_{max}<GM/({\cal R}R)$, the plasma continues its free fall down to the magnetosphere, still with the temperature $T_{max}$, with the
subsequent formation of a shock above the magnetosphere. However,
if $T_{max}> GM/({\cal R}R)$, settling accretion will work even for low wind velocities. 

For high-wind stellar velocities $v_w\gtrsim 100$~km/s, the post-shock temperature is higher than $T_{max}$, photoionization is unimportant, and the settling accretion regime 
is established if the radiation cooling time is longer than the accretion time. From a comparison of these timescales, we find the critical accretion rate as a function 
of of the wind velocity below which the settling accretion regime is possible:
\beq{}
\dot M_{16}^{\ddag}\lesssim 0.12 v_7^{3.2}\,.
\eeq
 
Here we stress the difference of the critical acccretion rate
$\dot M^{\ddag}$ and  
$\dot M^\dag$, derived earlier. For $\dot M>\dot M^{\ddag}$ he plasma rapidly cools down 
in the gravitational capture region and free-fall accretion begins (unless photoionization
heats up the plasma above the adiabatic value at some radius), while at  
$\dot M>\dot M^\dag
\simeq 4\times 10^{16}$~g/s, determined by \Eq{M*} a free-fall gap appears immediately above the 
neutron star magnetosphere.

\subsection{On the possibility of the propeller regime}
\label{s:propeller}

The very slow rotation of the neutron stars in X-ray pulsars considered here 
(GX 1+4, GX 301-2, Vela X-1, SXP 1062, 4U 2204+56) with 
$\omega^* (R_A)<\omega_K(R_A)$ (see Table 1)  
makes it hard for these sources to enter the propeller regime 
where matter is ejected with parabolic velocities from the magnetosphere and the neutron star spins down.
 
Let us therefore start with estimating the important 
ratio of viscous tensions ($\sim B_tB_p$) 
to the gas pressure ($\sim B_p^2$) at the magnetospheric boundary. This ratio is proportional to 
$B_t/B_p$ (see \Eq{BtBpnum}) and is always much smaller than 1 (see Table 1),
i.e. only large-scale convective  
motions where the characteristic hierarchy of 
eddies scales with radius can be established in the shell. 

When $\omega^*>\omega_K(R_A)$, a centrifugal barrier is formed and accretion stops (the propeller regime). In that case the maximum possible braking torque is $\sim -\mu^2/R_A^3$
due to the strong coupling between the plasma and the magnetic field. Note that in the propeller state, interaction of the plasma with the magnetic field is by strong coupling, i.e. the toroidal 
magnetic field component $B_t$ is comparable to the poloidal one $B_p$. 
It can not be excluded that a hot iso-angular-momentum envelope could exist 
in this case as well, which would then remove angular momentum from the rotating magnetosphere. If the characteristic cooling time of the gas in the envelope is short
in comparison to the falling time of matter, the shell disappears and 
one can expect the formation of a `storaging' thin Keplerian disc around the 
neutron star magnetosphere \cite{SunyaevShakura77}. There is no accretion of matter through such a disc. It 
only serves to remove angular momentum from the magnetosphere.

\subsection{Effects of the hot shell on the X-ray energy and power spectrum}

The spectra of X-ray pulsars are dominated by emission generated in 
the accretion column. The hot optically thin shell produces its own thermal emission, but even if 
all gravitational energy were released in the shell, the ratio of the X-ray luminosity 
from the shell to that of the accretion column would be about the ratio of the
magnetospheric radius to that of the neutron star, i.e. one percent or less. In reality, the luminosity from the shell is much smaller. The shell should scatter X-ray radiation from the accretion column, but 
for this effect to be substantial, the Comptonization parameter $y$ must be of 
the order of one. The Thomson depth in the shell is, however, very small. Indeed, from the mass continuity equation 
and \Eq{RA} for the Alfv\'en radius and \Eq{fu} for the factor $f(u)$, we get:
$$
\tau_T=\int_{R_A}^{R_B}n_e(R)\sigma_T dR
\approx 3.2\times 10^{-3} \dot M_{16}^{8/11}\mu_{30}^{-2/11}\,.
$$
Therefore, for the characteristic temperatures near the magnetosphere (see [\Eq{hse_sol}]) the parameter $y$ is 
$$
y=\frac{4kT}{m_ec^2}\tau_T\approx 2.4\times 10^{-3}\,.
$$
This means that the X-ray spectrum, formed in the region of energy conversion close to the surface of the neutron star is not expected to be significantly altered by scattering in the hot shell.

Large-scale convective motions in the shell introduce 
an intrinsic time-scale of the order of
the free-fall time that could give rise to 
features (e.g. QPOs) in the power spectrum of variability. 
QPOs were reported in some X-ray pulsars (see \cite{Marykutty_ea10} and references therein). However, the expected frequencies of any QPOs arising in our model would be of the order of mHz, much higher than those reported.

A stronger effect could be the appearance of a dynamical instability in the shell due to increased Compton cooling
and hence increased mass accretion rate through the shell. This may lead to a complete collapse of the shell triggering an X-ray outburst with duration similar to the free-fall time scale of the shell ($\sim 1000$~s). Such transient behaviour is observed in the supergiant fast X-ray transients (SFXTs) see \cite{Ducci_ea10}.) 
The possible development of such a scenario depends on the specific parameters
of the shell and needs to be further investigated.

\subsection{Can accretion discs (prograde or retrograde) 
be present in these pulsars?}

Our analysis of the sample of pulsars in Section \ref{s:applications} suggested that in a convective shell an 
iso-angular-momentum distribution is the most plausible. Therefore,   
we shall below consider only this case, i.e. using the rotation law $\omega\sim R^{-2}$. 
As follows from \Eq{sd_eq1}, at $\dot \omega^*=0$ the equilibrium angular 
frequency of the neutron star is
\beq{equilib}
\omega^*_{eq}=\omega_B\frac{1}{1-z/Z}\myfrac{R_B}{R_A}^2\,.
\eeq
We stress that such an equilibrium in our model is possible only 
when a shell is present. At high accretion rates 
$\dot M>\dot M_*\simeq 4\times 10^{16}$~g/s accretion proceeds 
in the free-fall regime (with no shell present).

The equilibrium period for an X-ray pulsar in the quasi-spherical settling accretion regime can be derived using the formula \Eq{Peq}: 
$
P_{eq}\simeq 1300 [\hbox{s}]\mu_{30}^{12/11}(P_b/10\hbox{d})
\dot M_{16}^{-4/11}v_8^4\,.
$

For standard disc accretion, the equilibrium period is
\beq{P_eqd}
P_{eq,d}\approx 7[\hbox{s}] \mu_{30}^{6/7}\dot M_{16}^{-3/7}\,,
\eeq
and the long periods observed in some X-ray pulsars can thus, if a is disc present, be explained only
assuming a very high magnetic field of the neutron star. Retrograde accretion discs are also discussed in the literature (see e.g. \cite{Nelson_ea97} and references therein). Torque reversals produced by temporary forming retrograde discs can in principle lead to very long periods for X-ray pulsars even with standard magnetic fields. Such retrograde discs could be formed as a result of inhomogeneities in the captured stellar wind \cite{Ruffert97, Ruffert99}. The scenario could, in principle, work for pulsars at high accretion rate, too high for a hot envelop to form. 

In the case of GX 1+4, however, it is highly unlikely to observe a retrograde disk 
on a time scale much longer than the orbital period (see a more detailed discussion 
of this issue in \cite{GonzalezGalan_ea12}). For both GX 301-2 and Vela X-1, the observed positive
torque-luminosity correlation (see Figs. \ref{f:gx301} and \ref{f:velaX1}) also rules
out a retrograde disc in any of these systems.

To conclude this discussion section, we should mention that in reality, all pulsars (including those considered here) demonstrate a complex quasi-stationary behaviour with dips, outbursts, etc. 
These considerations are beyond the scope of this paper and definitely 
deserve further observational and theoretical studies.

\section{Conclusions}
\label{s:conclusions} 

In  \cite{Shakura_ea12} we presented a theoretical model for quasi-spherical 
subsonic accretion onto slowly rotating magnetized 
neutron stars. 
In this model the accreting matter is gravitationally 
captured from the stellar wind of the optical companion and subsonically 
settles down onto the rotating magnetosphere forming an extended quasi-static shell.
This shell mediates the angular momentum removal from the rotating neutron star 
magnetosphere by large-scale convective motions. Depending on the angular velocity of the rotating matter close to the magnetospheric boundary this type of accretion can cause the neutron star to either spin up or spin down. 

A detailed analysis and comparison with observations of the two X-ray pulsars GX 301-2 and Vela X-1, both demonstrating positive torque-luminosity correlations near the
equilibrium neutron star spin period, shows that the convective motions are most likely 
strongly anisotropic, and the rotational velocities in the shell $\omega\sim R^{-2}$ have a 
near iso-angular-momentum distribution.
We note that a statistical analysis of long-period X-ray pulsars with Be-components in SMC by
\cite{ChashkinaPopov12} also favored the rotation law $\omega\sim R^{-2}$.

The accretion rate through the shell is determined by the ability of the plasma to enter the magnetosphere. The settling regime of accretion which allows angular momentum removal
from the neutron star magnetosphere can be realized for moderate accretion rates 
$\dot M< \dot M_*\simeq 4\times 10^{16}$~g/s. At higher accretion rates a free-fall gap 
above the neutron star magnetosphere appears due to rapid Compton cooling, and accretion 
becomes highly non-stationary. 

From observations of the spin-up/spin-down rates (the angular rotation frequency derivative 
$\dot \omega^*$, or  $\partial\dot\omega^*/\partial\dot M$ near the torque reversal) 
of slowly rotating equilibrium X-ray pulsars  with known orbital periods it is possible 
to determine the main dimensionless parameters of the model, as well as to estimate the magnetic field of the neutron star. Such an analysis revealed a good agreement between 
magnetic field estimates obtained using our model and those derived from cyclotron 
line measurements for the pulsars 
GX 301-2 and Vela X-1.

Using measurements of the spin period and the orbital period together with an estimate of the neutron star magnetic field $\mu$, our model furthermore offers a possibility to estimate the stellar wind velocity of the companion, without the need for complicated spectroscopic measurements.

For non-equilibrium pulsars there is a maximum possible spin-down rate, depending on the spin period $P^*$, the orbital period $P_b$, the neutron star magnetic field $\mu$ and the wind velocity  $v_w$. For such pulsars it is possible to estimate a lower limit on the neutron star magnetic field using the observed spin-down rate and X-ray luminosity. For the pulsars GX 1+4, SXP 1062, 4U 2206+54 investigated here, our estimates are all in agreement with standard field values and observed cyclotron line measurements.   

In our model for quasi-spherical subsonic accretion, long-term spin-up/spin-down as observed in some X-ray pulsars can be quantitatively explained by a change in the mean mass accretion rate onto the neutron star (and the corresponding mean X-ray luminosity). Clearly, these changes are related to the stellar wind properties. 

The model also predicts the specific behaviour 
of the variations in $\delta \dot \omega^*$, observed on top of a steady spin-up or spin-down,
as a function of mass accretion rate fluctuations $\delta \dot M$. There is a critical accretion rate $\dot M_{cr}$ below which an anti-correlation of $\delta \dot \omega^*$ with $\delta \dot M$ should occur (the case of 
GX 1+4 at the steady spin-down state currently observed), and above which 
$\delta \dot\omega^*$ should correlate with $\delta \dot M$ fluctuations (the case 
of Vela X-1, GX 301-2, and GX 1+4 in the steady spin-up state). The model explains quantitatively the relative amplitude and the sign 
of the observed frequency fluctuations in GX 1+4.  

\appendix
\numberwithin{equation}{section}

\section{The structure of a quasi-spherical rotating shell with accretion}

\subsection{Basic equations}
Let us first write down the Navier-Stokes equations in spherical coordinates $R, \theta, \phi$. Due to the huge Reynolds numbers in the shell ($\sim 10^{15}-10^{16}$ for a 
typical accretion rate of $10^{17}$~g/s and magnetospheric radius $\sim 10^8$~cm), there must be strong turbulence.  
In this case the Navier-Stokes equations are usually called the Reynolds equations.
In the general case, the turbulent viscosity may depend on the coordinates,
so the equations take the form:
 
1. Mass continuity equation:
\beq{mass_cont}
   \frac{\partial \rho}{\partial t}+ 
\frac{1}{R^2}\frac{\partial}{\partial R}\left(R^2 \rho u_r\right)+ \frac{1}{R \sin\theta}\frac{\partial}{\partial \theta}\left(\sin\theta\, \rho u_\theta\right) + \frac{1}{R \sin\theta}\frac{\partial \rho u_\phi}{\partial \phi}  = 0. 
\eeq
 
2. The $R$-component of the momentum equation:
\beq{v_R}
\displaystyle\frac{\partial u_r}{\partial t} + u_r \displaystyle\frac{\partial u_r}{\partial R} + \displaystyle\frac{u_{\theta}}{R} \displaystyle\frac{\partial u_r}{\partial \theta}+ \displaystyle\frac{u_{\phi}}{R \sin\theta} \displaystyle\frac{\partial u_r}{\partial \phi}  - \displaystyle\frac{u_{\phi}^2 + u_{\theta}^2}{R} = 
-\displaystyle\frac{GM}{R^2}+N_R
\eeq
 
3. The $\theta$-component of the momentum equation:  
\beq{v_theta}    
\displaystyle\frac{\partial u_{\theta}}{\partial t} 
+ u_r \displaystyle\frac{\partial u_{\theta}}{\partial R}
+ \displaystyle\frac{u_{\theta}}{R} \displaystyle\frac{\partial u_{\theta}}{\partial \theta}  
+ \displaystyle\frac{u_{\phi}}{R \sin\theta} \displaystyle\frac{\partial u_{\theta}}{\partial \phi} 
+ \displaystyle\frac{u_r u_{\theta} - u_{\phi}^2 \cot\theta}{R} = 
N_\theta
\eeq
 
4. The $\phi$-component of the momentum equation:  
\beq{v_phi}
\displaystyle\frac{\partial u_{\phi}}{\partial t} 
+ u_r \displaystyle\frac{\partial u_{\phi}}{\partial R} 
+ \displaystyle\frac{u_{\theta}}{R} \displaystyle\frac{\partial u_{\phi}}{\partial \theta} 
+ \displaystyle\frac{u_{\phi}}{R \sin\theta} \displaystyle\frac{\partial u_{\phi}}{\partial \phi} 
+ \displaystyle\frac{u_r u_{\phi} + u_{\phi} u_{\theta} \cot\theta}{R} =
N_\phi  
\eeq

Here the force components (including viscous force and gas pressure gradients) read:
\beq{N_R}
\rho N_R=\frac{1}{R^2}\frac{\partial}{\partial R}\left( R^2 W_{RR}\right)
+\frac{1}{\sin\theta\,R}\frac{\partial}{\partial \theta}\left(  W_{R\theta}\sin\theta\right)
+\frac{1}{\sin\theta\,R}\frac{\partial}{\partial \phi}W_{R\phi}-\frac{W_{\theta\theta}}{R}
-\frac{W_{\phi\phi}}{R}
\eeq
 
\beq{N_theta}
\rho N_\theta=\frac{1}{R^2}\frac{\partial}{\partial R}\left( R^2 W_{\theta R}\right)
+\frac{1}{\sin\theta\,R}\frac{\partial}{\partial \theta}\left(  W_{\theta\theta}\sin\theta\right)
+\frac{1}{\sin\theta\,R}\frac{\partial}{\partial \phi}W_{\theta\phi}-\cot\theta\frac{W_{\theta\theta}}{R}
\eeq
 
\beq{N_phi}
\rho N_\phi=\frac{1}{R^3}\frac{\partial}{\partial R}\left( R^3 W_{\phi R}\right)
+\frac{1}{\sin\theta\,R}\frac{\partial}{\partial \theta}\left(  W_{\phi\theta}\sin\theta\right)
+\frac{1}{\sin\theta\,R}\frac{\partial}{\partial \phi}W_{\phi\phi}
\eeq  
 
The components of the stress tensor include a contribution from both the gas pressure $P_g$ 
(assumed to be isotropic) and the turbulent pressure $P^t$ (generally anisotropic). In their definition we shall follow 
the classical treatment by Landau and Lifshitz \cite{LandauLifshitz86} but with the inclusion of 
anisotropic turbulent pressure:
 
\beq{}
W_{RR}=-P_g-P_{RR}^t+2\rho\nu_t\frac{\partial u_r}{\partial R}-\frac{2}{3}\rho\nu_t{\rm div} {\bf u}
\eeq
 
\beq{}
W_{\theta\theta}=-P_g-P_{\theta\theta}^t+2\rho\nu_t\left(\frac{1}{R}\frac{\partial u_\theta}{\partial \theta}
+\frac{u_r}{R}\right)-\frac{2}{3}\rho\nu_t{\rm div} {\bf u}
\eeq
 
\beq{}
W_{\phi\phi}=-P_g-P_{\phi\phi}^t+
2\rho\nu_t\left(\frac{1}{R\sin\theta}\frac{\partial u_\phi}{\partial \phi}
+\frac{u_r}{R}+\frac{u_\theta\cot\theta}{R}\right)-\frac{2}{3}\rho\nu_t{\rm div} {\bf u}
\eeq
 
\beq{}
W_{R\theta}=\rho\nu_t\left(\frac{1}{R}\frac{\partial u_r}{\partial \theta}
+\frac{\partial u_\theta}{\partial R}-\frac{u_\theta}{R}\right)
\eeq

\beq{}
W_{\theta\phi}=\rho\nu_t\left(\frac{1}{R\sin\theta}\frac{\partial u_\theta}{\partial \phi}
+\frac{1}{R}\frac{\partial u_\phi}{\partial\theta}-\frac{u_\phi\cot\theta}{R}\right)
\eeq
 
\beq{}
W_{R\phi}=\rho\nu_t\left(\frac{1}{R\sin\theta}\frac{\partial u_r}{\partial \phi}
+\frac{\partial u_\phi}{\partial R}-\frac{u_\phi}{R}\right)
\eeq
In our problem the anistropy of the turbulence is such that
$P_{RR}^t=P_\parallel^t$, 
$P_{\theta\theta}^t=P_{\phi\phi}^t=P_\perp^t$. The turbulent pressure components can be 
expressed through turbulent Mach numbers and will be given in Appendix E.
  
${\rm div}{\bf u}$ in spherical coordinates is:
\beq{}
{\rm div}{\bf u}=
\frac{1}{R^2}\frac{\partial}{\partial R}\left(R^2 u_r\right)
+ \frac{1}{R \sin\theta}\frac{\partial}{\partial \theta}\left(\sin\theta\, u_\theta\right) + \frac{1}{R \sin\theta}\frac{\partial u_\phi}{\partial \phi}. 
\eeq

\subsection{Symmetries of the problem} 

We shall consider axially-symmetric ($\displaystyle\frac{\partial}{\partial \phi}=0$), 
stationary ($\displaystyle\frac{\partial}{\partial t}=0$), and only radial accretion
 ($u_\theta=0$). Under these conditions, from the continuity equation \Eq{mass_cont} we obtain:
\beq{dotM}
\dot M= 4\pi R^2\rho u_R=const\,.
\eeq
The constant here is determined from the condition of plasma leakage through the magnetosphere.

Let us rewrite the Reynolds equations under the above assumptions.
The $R$-component of the momentum \Eq{v_R} equation becomes:
\beq{v_R1}
 \rho\left(u_R \displaystyle\frac{\partial u_R}{\partial R} 
  - \displaystyle\frac{u_{\phi}^2}{R}\right) 
= 
-\rho \displaystyle\frac{GM}{R^2}+
\frac{1}{R^2}\frac{\partial}{\partial R}\left( R^2 W_{RR}\right)
+\frac{1}{\sin\theta\,R}\frac{\partial}{\partial \theta}\left(  W_{R\theta}\sin\theta\right)
-\frac{W_{\theta\theta}}{R}
-\frac{W_{\phi\phi}}{R}
\eeq
The $\theta$-component of the momentum equation: 
\beq{v_theta1}    
-\rho\frac{ u_{\phi}^2 \cot\theta}{R} 
=
\frac{1}{R^2}\frac{\partial}{\partial R}\left( R^2 W_{\theta R}\right)
+\frac{1}{\sin\theta\,R}\frac{\partial}{\partial \theta}\left(  W_{\theta\theta}\sin\theta\right)
-\cot\theta\frac{W_{\theta\theta}}{R}
\eeq
The $\phi$-component of the momentum equation: 
\beq{v_phi1}
\rho\left(u_R \displaystyle\frac{\partial u_{\phi}}{\partial R} 
+ \displaystyle\frac{u_R u_{\phi}}{R}\right) 
=\frac{1}{R^3}\frac{\partial}{\partial R}\left( R^3 W_{\phi R}\right)
+\frac{1}{\sin\theta\,R}\frac{\partial}{\partial \theta}\left(  W_{\phi\theta}\sin\theta\right)
\eeq

The components of the stress tensor with anisotropic turbulence 
take the form:

\beq{W_RR}
W_{RR}=-P_g-P_{\parallel}^t-\frac{4}{3}\rho\nu_t\left(\frac{u_r}{R}-\frac{\partial u_r}{\partial R}\right)
\eeq
 
\beq{W_tt}
W_{\theta\theta}=-P_g-P_{\perp}^t+\frac{2}{3}\rho\nu_t\left(\frac{u_r}{R}-\frac{\partial
 u_r}{\partial R}\right)
\eeq
 
\beq{W_pp}
W_{\phi\phi}=-P_g-P_{\perp}^t+\frac{2}{3}\rho\nu_t\left(\frac{u_r}{R}-\frac{\partial u_r}{\partial R}\right)
\eeq
 
\beq{W_rt}
W_{R\theta}=\rho\nu_t\frac{1}{R}\frac{\partial u_r}{\partial \theta}
\eeq
 
\beq{W_tf}
W_{\theta\phi}=\rho\nu_t\left(
\frac{1}{R}\frac{\partial u_\phi}{\partial\theta}-\frac{u_\phi\cot\theta}{R}\right)
\eeq
 
\beq{W_rf}
W_{R\phi}=\rho\nu_t\left(\frac{\partial u_\phi}{\partial R}-\frac{u_\phi}{R}\right)
\eeq 

The main problem in describing gas dynamic flows with turbulence is in finding the kinematic viscosity parameter $\nu_t$. As is well known, in the case of laminar flows the viscosity parameter $\nu$ is dependent only on the properties of the medium (liquid or gas). When turbulence is present, however, this coefficient is determined also by the macroscopic properties of the flow itself. The are some empirical relations which in principle can be verified experimentally. Most often the so called turbulent mixing length $l_t$ is introduced. Furthermore L. Prandtl in his works        introduced for plane-parallel shear flows (along the x-axis to be specific) the relation between the turbulent mixing length $l_t$, the velocity of the turbulent flow $u_t$ and the characteristic amount of shear in the direction perpendicular to the average flow ($z$):
\beq{}
\nu_t=C_0 l_t \left|\frac{du}{dz}\right|
\eeq
where $C_0\sim 1$ is a universal dimensionless constant, the exact numerical value of which should be determined from a theory that currently does not exist. In this way, the dependence of the turbulent stresses on the shear value becomes quadratic:
$$
W_{zx}=\rho C_0 \myfrac{du}{dz}^2\,,
$$
and a non-linearity is formed which in the general case makes the problem a lot more difficult.

First, we consider the general Prandtl law for turbulent viscosity in the case of an axi-symmetric flow. In the case of strong anisotropy there is one more empirical law for describing the turbulent viscosity, the Wasiuty\'nski-law (see below), which is not reduced to the Prandtl law in the case of isotropic turbulence. This more general case for anisotropic turbulence will be discussed separately in Appendix C. 

\section{Structure of the shell in the case of turbulent viscosity according to the Prandtl law}

\subsection{The empirical Prandtl law for axisymmetric flows with turbulent viscosity}. 
We consider an axisymmetric flow with a very large Reynolds number. By generalizing the 
Prandtl law for the turbulent velocity obtained for plane parallel flows, 
the turbulent velocity scales as $u_t\sim l_t R(\partial\omega /\partial R)$. From the similarity laws 
of gas-dynamics we assume $l_t\sim R$,  so
\beq{u_t}
u_t=C_1R^2\left|\frac{\partial\omega }{\partial R}\right|\,.
\eeq

We note that in our case the turbulent velocity is determined by convection, and thus $u_t\lesssim 0.5 u_{ff}$ (see Appendix D). This implies that the constant $C_1$ scales as 
\beq{}
C_1\sim u_t/\langle u_\phi \rangle, 
\eeq
and can be very large since $\langle u_\phi \rangle \ll u_t$. 

The turbulent viscosity coefficient thus reads:
\beq{nut}
\nu_t=\langle u_t l_t\rangle=C_2C_1 R^3\left|\frac{\partial \omega}{\partial R}\right|
\eeq
Here $C_2\approx 1/3$ is a numerical factor originating from statistical averaging.
Below we shall combine $C_1$ and $C_2$ into the new coefficient $C=C_1C_2$, 
which can be much larger than unity. 

For such a viscosity prescription the turbulent stresses $W_{R\phi}$ are equal to
\beq{WRphi_P}
W_{R\phi}=\rho \nu_t R \frac{\partial \omega}{\partial R}=\rho C R^4\myfrac{\partial \omega}{\partial R}^2.
\eeq 

\subsection{The angular momentum transport equation}
A similar problem (that of a rotating sphere in a viscous fluid) was solved in Landau and Lifshitz \cite{LandauLifshitz86}. They showed that the variables here become separated and 
$u_{\phi}(R,\theta)= u_\phi(R)\sin\theta$. Note that the angular velocity 
$\omega(R)=u_\phi(R)/R$ is independent of the polar angle $\theta$. 
Our problem is different from that of the sphere in a viscous fluid in several respects: 1) there is a force of gravity present, 2) the turbulent viscosity varies with $R$ and can 
in principle depend on $\theta$, and 3) there is radial motion of matter (accretion). These differences lead, 
as will be shown below, to the radial dependence 
$u_\phi (R)\propto R^{-1/2}$. 
(We recall that for a rotating sphere in a viscous fluid $u_\phi\propto R^{-2}$). 

Let us start with solving \Eq{v_phi1}. First, we note that for 
$u_\phi(\theta)\sim \sin\theta$, according to \Eq{W_tf}, $W_{\theta\phi}=0$. 
Further, making use of the continuity equation \Eq{dotM} and the definition of angular velocity, 
we rewrite \Eq{v_phi1} in the form of angular momentum transfer by viscous forces:
\beq{amt}
\sin\theta\frac{\dot M}{R}\frac{\partial}{\partial R}\omega R^2 = \frac{4\pi}{R}\frac{\partial}{\partial R}R^3W_{R\phi}\,.
\eeq
 
We rewrite equation \Eq{W_rf} using the derivative of the angular velocity:
\beq{W_rf1}
W_{R\phi}=\rho \nu_t R \frac{\partial \omega}{\partial R}\sin\theta\,.
\eeq
Substituting this expression into  \Eq{amt} and integrating over $R$, we get
\beq{amt1}
\dot M \omega R^2= 4\pi \rho \nu_t R^4 \frac{\partial \omega}{\partial R}+D\,,
\eeq
where $D$ is an integration constant. 
This equation for angular mometum transport by turbulent viscosity is similar to that 
of disc accretion \cite{ShakuraSunyaev73}, but different due to spherical symmetry of our problem. 

The left part of  \Eq{amt1} is simply advection of specific 
angular momentum averaged over the sphere ($1/2\int_0^\pi\omega R^2 \sin^2\theta \sin\theta d\theta=1/3 \omega R^2$) by the average motion toward the gravitational center (accretion). $\dot M$ is negative 
as well as $\frac{\partial \omega}{\partial R}$. 
The first term on the right describes transport of angular momentum outwards by turbulent viscous forces. 

The constant $D$ is determined from the equation
\beq{D1}
D=\myfrac{K_1}{\zeta}K_2 \frac{\mu^2}{R_A^3}\frac{\omega_m-\omega^*}{\omega_K(R_A)}
\eeq
(see \Eq{sd1} in the text). We consider accretion onto a magnetized neutron star. 
 When $D<0$, the advection term in the left part of \Eq{amt1} 
dominates over viscous angular momentum transfer outwards. Oppositely,
when $D>0$, the viscous term in the right part of \Eq{amt1} dominates. 
In the case of $\dot M=0$ (no plasma enters the magnetosphere), 
there is only angular momentum transport outwards by viscous forces.

Now let us rewrite \Eq{D1} in the form
\beq{D2}
D=\myfrac{K_1}{\zeta}K_2
\frac{\mu^2}{R_A^6}R_A^3\frac{\omega_m-\omega^*}{\omega_K(R_A)}
\eeq
and use the pressure balance condition
\beq{}
P(R_A)=P_g(R_A)(1+\gamma m_t^2)=
\frac{B^2(R_A)}{8\pi}=
\frac{K_2}{2\pi}\frac{\mu^2}{R_A^6}\,.
\eeq
Using the mass conitnuity equation in the form 
\[
|\dot M|=4\pi R^2\rho f(u)\sqrt{GM/R}\,,
\]
and the expression for the gas pressure \Eq{P(RA)}, 
we write the integration constant $D/|\dot M|$ in the form
\beq{}
\frac{D}{|\dot M|}=\myfrac{K_1}{\zeta}\frac{(\gamma-1)}{\gamma}\psi(\gamma, m_t)
\frac{(\omega_m-\omega^*)R_A^2}{2\sqrt{2}f(u)}(1+\gamma m_t^2)\,.
\eeq  

Let us consider the case when the neutron star rotates close to equilibrium $\dot \omega^*=0$. In this case according to 
\Eq{sd_eq} 
\beq{do}
\omega_m-\omega^*=-\frac{z}{Z}\omega^*\,,
\eeq
and thus using definition of $Z$ [\Eq{Zdef}], we obtain:
\beq{Deq}
\frac{D}{|\dot M|}=-zR_A^2 \omega^*\,.
\eeq
We would like to stress that the value of the constant $D$ is fully determined by 
the dimensionless specific angular momentum of matter at the Alfv\'en radius $z$. 

\subsection{The angular rotation law inside the shell}

Let us now use \Eq{amt1} to find the rotation law $\omega(R)$. At large distances $R\gg R_A$ (we would like to remind the reader that $R_A$ is the bottom radius of the shell),
the constant $D$ is small relative to the other terms, so we can set $D\approx 0$.
Thus, to obtain the rotation law we shall neglect this constant in the right part
of \Eq{amt1}. Next, we substitute \Eq{nut} 
and make use of
the solution for the density (which, as we shall show below, remains the same as in the hydrostatic solution)  
\beq{rho_R}
\rho(R)=\rho(R_A)\myfrac{R_A}{R}^{3/2}\,
\eeq
in equation \Eq{amt1} to obtain:
\beq{amt2}
\left|\dot M\right| \omega R^2= 4\pi \rho(R_A) 
\myfrac{R_A}{R}^{3/2}CR^7 \myfrac{\partial \omega}{\partial R}^2\,.
\eeq 
After integrating this equation, we find
\beq{sqrtomega}
2\omega^{1/2}=\pm \frac{4}{3}\frac{K^{1/2}}{R^{3/4}}+D_1\,,
\eeq
where
\beq{}
K=\frac{|\dot M|}{4\pi \rho(R_A) C R_A^{3/2}}
\eeq
and $D_1$ is some integration constant. In \Eq{sqrtomega} we use only the positive 
solution (the minus sign with constant $D_1>0$ would correspond to 
a solution with the angular velocity growing outwards, which is possible if
the pulsar has a very long spin period, i.e. almost does not rotate at all). If $D_1\ne 0$, at large $R\gg R_A$ (in the zone close to the bow shock) the solid
body rotation law would lead to $\omega\to const\approx \omega_B$. (However, we remind the reader
that our discussion is not applicable close to the bow shock region.) At small distances from the Alfv\'enic surface the effect of this constant is small and we shall neglect it in the calculations below. Then we find 
\beq{omega}
\omega(R)=\frac{4}{9}\frac{|\dot M|}{4\pi \rho(R_A)CR_A^3}\myfrac{R_A}{R}^{3/2}
\eeq 
i.e. the quasi-Keplerian law $\omega(R)=\omega_m(R_A/R)^{3/2}$. 
The value $\omega_m$ in the solution given by \Eq{omega}
is obtained after substituting $\dot M$ from the continuity equation at $R=R_A$ into 
\Eq{omega}:
\beq{omega_m}
\omega_m\equiv
 \tilde \omega \omega(R_A)=\frac{4}{9}\tilde \omega \frac{|u_r(R_A)|}{CR_A}\,.
\eeq
(Here we have introduced the correction factor $\tilde\omega>1$ to account for the deviation of the exact solution from the Keplerian law 
close to $R_A$).

As $u_R(R_A)$ is smaller than the free-fall velocity, the above formula
implies that $\omega_m< \omega_K(R_A)$, lower than 
the Keplerian angular frequency. For self-consistency
the coefficient $C$ in the Prandtl law is determined, according to 
\Eq{omega_m}, by the 
ratio of the radial velocity $u_R$ to the rotational velocity of matter $u_\phi$: 
\beq{C1}
C=\frac{4}{9}\tilde\omega \frac{|u_r(R_A)|}{\omega_mR_A}
=\frac{4}{9}\tilde \omega \frac{|u_r(R_A)|}{u_\phi(R_A)}\,.
\eeq 
We note that this ratio is independent of 
the radius $R$ and is actually constant across the shell. Indeed, 
the radial dependence of the velocity $u_R$ 
follows from the continuity equation with account for 
the density distribution \Eq{rho_R}
\beq{u_rR}
u_r(R)=u_r(R_A)\myfrac{R_A}{R}^{1/2}\,.
\eeq
For a quasi-Keplerian law
$u_\phi(R)\sim 1/R^{1/2}$, so the ratio $u_r/u_\phi$ is constant.

Finally, 
the angular frequency of the rotation of the shell near the magnetosphere $\omega_m$ 
is related to the angular frequency of the motion of matter near the bow-shock as
\beq{}
\omega_m=\tilde\omega\omega_B\myfrac{R_B}{R_A}^{3/2}\,.
\eeq
In fact, when approaching $R_A$, the integration
constant $D$ (which we neglected at large distances $R\gg R_A$) should be taken into account.
The rotational law will thus somewhat differ from a quasi-Keplerian close to the magnetosphere. 

We stress the principal difference between this regime of accretion
and disc accretion. For disc accretion the radial velocity is much smaller than the
turbulent velocity, and the tangential velocity is almost Keplerian and is much larger than the
turbulent velocity.
The radial velocity in the quasi-spherical case is not determined by
the rate of the angular momentum removal. It is determined only by the 
"permeability" of the neutron star magnetosphere for infalling matter.  
In our case we assume that the radial velocity is of the order of the velocity of convective motions in the shell. The tangential velocity for the obtained quasi-Keplerian law is much smaller than the velocity of 
the convective motions. Note also that in the case of disc accretion the turbulence can be parametrized by only one dimensionless parameter $\alpha\approx u_t^2/u_s^2$ with 
$0 < \alpha <1$ 
\cite{ShakuraSunyaev73}. The matter in an accretion disc rotates differentially
with a supersonic (almost Keplerian) velocity, while in our case the shell rotates differentially
with a clearly subsonic velocity at any radius, and the turbulence in the shell is essentially subsonic. 
Also, our case with an extended shell is of course strongly different from the regime of freely falling matter
with a standing shock above the magnetosphere \cite{AronsLea76a}.

\subsection{The case without accretion}

Now let us consider the case where the plasma can not enter the magnetosphere
and no accretion onto the neutron star occurs. 
This case is similar to the subsonic propeller regime considered by Davies and Pringle \cite{DaviesPringle81}. \Eq{amt1} then takes the form:
\beq{amt01}
0=4\pi \rho \nu_t R^4 \frac{\partial \omega}{\partial R}+D\,.
\eeq 
(Remember that the constant $D$ is determined by the spin-down
rate of the neutron star, $D=I\dot \omega^*<0$).
Solving this equation as above, we find for the rotation law without accretion:
\beq{omega0}
\omega(R)=\omega_m\myfrac{R_A}{R}^{7/4}\,,
\eeq
where
\beq{}
\omega_m=\frac{I|\dot \omega^*|}{7\pi\rho(R_A)\nu_t(R_A)R_A^3}\,.
\eeq
From \Eq{nut} we find
\beq{nut1}
\nu_t(R_A)=\frac{7}{4}C\omega_mR_A^2\,,
\eeq
and thus for 
$\omega_m$ we obtain:
\beq{A45}
\omega_m=\frac{2}{7}\myfrac{I|\dot \omega^*|}{\pi C\rho(R_A)R_A^5}^{1/2}\,.
\eeq
However, $\omega_m$ is also related to the bow-shock region parameters as
\beq{}
\omega_m=\omega_B\myfrac{R_B}{R_A}^{7/4}\,,
\eeq
which can in principle be used to further study this case, which we shall not do here.

\section{Structure of the shell and angular rotation law in case of turbulent viscosity according to Wasiuty\'nski}
\label{appC} 

Prandtl's law for viscosity that was used above relates the scale and velocity of turbulent pulsations with the average angular velocity and is commonly used when the turbulence is generated by the shear itself. In our problem, the turbulence is initiated by large-scale convective motions in the gravitational field. Due to convection, strong anisotropic
turbulent motions may appear (the radial dispersion of chaotic motions could be much larger than the dispersion in the tangential direction), and Prandtl's law may thus be inapplicable. 

Anisotropic turbulence is much more complicated and remains poorly studied. As a first step, we may adopt the empirical law for $W_{R\phi}$ as suggested 
by Wasiuty\'nski \cite{Wasiutinski46}:
\beq{aW1}
W_{R\phi}=2\rho(-\nu_t+\nu_r)\omega +\nu_r\rho R\frac{d\omega}{dR}\,,
\eeq
where the radial and tangential kinematic viscosity coefficients are 
\[
\nu_r=C_\parallel\langle |u_\parallel^t|\rangle R
\] 
\[
\nu_t=C_\perp\langle |u_\perp^t|\rangle R
\] 
respectively. The dimensionless constants $C_\parallel$ and $C_\perp$ are of the order of one. In the isotropic case $\nu_r=\nu_t$, $W_{R\phi}\sim d\omega/dR$, and in the strongly 
anisotropic case $\nu_r\gg\nu_t$, $W_{R\phi}\sim d(\omega R^2)/dR$. Using these definitions, let us substitute \Eq{aW1} into \Eq{amt}, and after integration over $R$ 
rewrite the latter in the form:
\beq{}
\omega R^2\left(1-\frac{2C_\perp\langle |u_\perp^t|\rangle}{|u_r|}\right)=
C_\parallel\frac{\langle |u_\parallel^t|\rangle }{|u_r|}\frac{Rd(\omega R^2)}{dR}-\frac{D}{|\dot M|}\,.
\eeq 

We note that due to self-similarity in the shell structure $u^t_\parallel\sim 
u^t_\perp\sim u_r\sim R^{-1/2}$, and thus the ratios $\langle |u_\parallel^t|\rangle/u_r$
and  $\langle |u_\perp^t|\rangle/u_r$ are constant. In this case the obvious solution to the above equation reads:
\beq{}
\omega R^2+\frac{D}{|\dot M|}\frac{1}{1-2C_\perp\frac{\langle |u_\perp^t|\rangle}{|u_r|}}=
\left[\omega_BR_B^2+\frac{D}{|\dot M|}\frac{1}{1-2C_\perp\frac{\langle |u_\perp^t|\rangle}{|u_r|}}\right]\myfrac{R_B}{R}^{\frac{|u_r|}{C_\parallel\langle |u_\parallel^t|\rangle}\left(1-2C_\perp\frac{\langle |u_\perp^t|\rangle}{|u_r|}\right)}
\eeq
(here the integration constant is defined as such that  $\omega(R_B)=\omega_B$).
 
Now let us consider the equilibrium situation where
$\dot\omega^*=0$. In this case, as we remember, 
\[
\frac{D}{|\dot M|}=-z\omega^* R_A^2\;,
\omega_m=(1-z/Z)\omega^*\,.
\] 

First, let us consider the case of strongly anisotropic, almost radial turbulence where $\langle |u_\perp^t|\rangle=0$. In this case, the specific angular momentum at the Alfv\'en radius is 
\beq{case1}
\omega_m R_A^2\left[1+\frac{z}{1-z/Z}\left(\myfrac{R_B}{R_A}^
\frac{|u_r|}{C_\parallel\langle|u^t_\parallel|\rangle}-1\right)\right]=
\omega_BR_B^2\myfrac{R_B}{R_A}^\frac{|u_r|}{C_\parallel\langle|u^t_\parallel|\rangle}\,.
\eeq
From this we see that in the case of very weak accretion (or, in the limit, when there is no accretion through the magnetosphere at all), $|u_R|\ll C_\parallel\langle|u^t_\parallel|\rangle$, and an almost iso-angular-momentum distribution of rotational velocities in the shell is formed.
 
The next case is where the amount of anisotropy is such that  
$C_\perp\langle|u^t_\perp|\rangle/|u_r|=1/2$. Then we have a strict iso-angular-momentum distribution in the shell:
 $\omega_mR_A^2=\omega_BR_B^2$. 
 
If the turbulence is fully isotropic $C_\perp\langle|u^t_\perp|\rangle=\
C_\parallel\langle|u^t_\parallel|\rangle=\tilde C\langle|u^t|\rangle$. Denoting $\epsilon=|u_r|/(\tilde C\langle |u^t|\rangle)$,
we find:
\beq{case2}
\omega_mR_A^2\left[1+\myfrac{z}{1-z/Z}\myfrac{1}{2/\epsilon-1}
\left(1-\myfrac{R_A}{R_B}^{2-\epsilon}\right)
\right]=\omega_BR_B^2\myfrac{R_A}{R_B}^{2-\epsilon}\,.
\eeq
Note that if $\epsilon\to 0$ (and there is no accretion through the magnetosphere), $\omega_m\to \omega_B$, 
and we have solid-body rotation without accretion (cf. the first case above!). For $\epsilon=3/2$, a near quasi-Keplerian angular rotation distribution may be established. We remind the reader that a similar quasi-Keplerian distribution was obtained in Appendix B above with the use of the Prandtl law for isotropic turbulent viscosity. In that case, this was the only solution. Here, in contrast, the quasi-Keplerian law is only one particular case of the general solution obtained using Wasiuty\'nsky's prescription for anisotropic turbulent viscosity.

As we have shown in the main text, a quasi-Keplerian rotation law 
is not favored by observations. We therefore conclude that the most likely velocity distribution in the shell is the near iso-angular-momentum one with anisotropic turbulence initiated by convection. Note that for thin accretion discs where the vertical height limits the scale of the turbulence, the Prandlt law for viscosity works very well
\cite{ShakuraSunyaev73}.

\section{Corrections to the radial temperature gradient}
 
Here we shall estimate how the radial temperature 
gradient differs from the adiabiatic law due
to the convective motions in the shell.
By multiplying \Eq{sd_om} by $(1/2)(\omega_m-\omega^*)$, we obtain the
convective heating rate 
caused by interaction of the shell with the magnetosphere:
 \beq{}
L_c=\frac{1}{2}Z\dot M R_A^2(\omega_m-\omega^*)^2\,.
\eeq
Multiplying the same 
\Eq{sd_om} with $\omega^*$ yields the rate of change of the mechanical energy of the neutron star
 \beq{}
L_k=Z\dot M R_A^2\omega^*(\omega_m-\omega^*)\,.
\eeq
The total energy balance is then
\beq{}
L_t=L_c+L_k=\frac{1}{2}Z \dot M R_A^2 (\omega_m^2-\omega^{*2})\,.
\eeq
Note that the obtained formula for $L_c$ is similar to that describing
energy release in the boundary layer of an accretion disc, see \cite{ShakuraSunyaev88}.  

The convective energy flux is:
\beq{q_m}
q_c=\frac{L_c}{4\pi R^2}=\frac{Z\dot M R_A^2(\omega_m-\omega^*)^2}{8\pi R^2}\,.
\eeq 

The convective energy flux can also be related to the entropy gradient as \cite{Shakura_ea77}):
\beq{8}
q_c=-\rho\nu_c T\frac{dS}{dR}\,,
\eeq
where $S$ is the specific entropy (per gram). Here $\nu_c$ 
is the radial turbulent heat conductivity 
\beq{}
\nu_c=<u_c l_c>=C_hu_cR\,,
\eeq
where the characteristic scale of convection $l_c\sim R$, the velocity of convective motions $u_c\sim c_s\sim R^{-1/2}$, and $C_h$ is a numerical coefficient of the order of one. Thus 
\beq{}
\nu_c=\nu_c(R_A)\myfrac{R}{R_A}^{1/2}\,.
\eeq 

Next, we make use of the thermodynamic identity for the specific enthalpy $H$:
\beq{tdid}
\frac{dH}{dR}=\frac{1}{\rho}\frac{dP_g}{dR}+T\frac{dS}{dR}\,.
\eeq
We remind the reader that the enthalpy can be written as 
\[
dH=c_p dT\,,
\]
where 
\[
c_p=T\myfrac{\partial S}{\partial T}_p=\frac{\gamma}{\gamma-1}\frac{{\cal R}}{\mu_m}
\]
is the specific heat capacity at constant pressure. 
Expressing $T(dS/dR)$ from 
\Eq{8} and making use of the hydrostatic equation [\Eq{hse_sol}] written as
\[
\frac{dP_g/\rho}{dR}=-\frac{{\cal R}}{\mu_m c_p}\frac{GM}{R^2}\psi(\gamma,m_t)\,.
\]
the thermodynamic identity \Eq{tdid} can be rewritten in the form 
\beq{tdid1}
\frac{dT}{dR}=-\frac{1}{c_p}\left[\frac{GM}{R^2}\psi(\gamma,m_t)
-\frac{Zu_r(R_A)}{2\nu_c(R_A)}\myfrac{R_A}{R}R_A^2(\omega_m-\omega^*)^2\right]\,.
\eeq 
By definition the adiabatic temperature gradient is determined by the first term
on the right hand side $(dT/dR)_{ad} =g/c_p$. 
Equation (\ref{tdid1}) can be integrated to find the actual dependence of the temperature on the radius in the convective shell:
\beq{Treal}
T=\frac{1}{c_p}\left[
\frac{GM}{R}\psi(\gamma,m_t)
-\frac{Zu_r(R_A)}{2\nu_c(R_A)}R_A^3(\omega_m-\omega^*)^2\ln\myfrac{R}{R_A}
\right]\,.
\eeq

Close to equilibrium ($I\dot \omega^*=0$), we can use \Eq{do} and write
\beq{}
T=\frac{1}{c_p}\left[\frac{GM}{R}\psi(\gamma,m_t)
-\frac{u_r(R_A)}{2C_h u_c(R_A)}\omega^{*2}R_A^2\frac{z^2}{Z}\ln\myfrac{R}{R_A}\right]\,.
\eeq
This solution shows that in the whole region between $R_A$ 
and $R_B$, for slowly rotating pulsars (i.e.,
in which $\omega_m\ll\omega_K(R_A)$), the temperature distribution is close to the adiabatic law with a
temperature gradient close to the adiabatic one [\Eq{hse_sol}]:
\beq{}
T\approx \frac{\gamma-1}{\gamma}\frac{GM}{{\cal R}R}\psi(\gamma,m_t)\,.
\eeq
Here we have only taken into account energy release due to the frequency difference 
near the magnetosphere. In reality, there may be additional sources of energy in the shell (e.g. the heat release during magnetic reconnection and turbulence (see Appendix E), etc.). 

\section{Dynamics of a stationary spherically-symmetric ideal gas flow}

In this Appendix, we write down the gas-dynamic equations of a 
spherically symmetric ideal gas flow onto a Newtonian gravitating center.
This problem was considered in the classical paper by Bondi (\cite{Bondi52}) for an adibatic accretion 
flow. Adiabatic gas outflows (stellar winds) were studied by Parker \cite{Parker63}. A thorough
and comprehensible discussion of such flows can be found in the monograph by V. Beskin \cite{Beskin05}. Here we focus on 
the role of the cooling/heating processes near the Alvenic surface, and also take into account the effects of turbulence and/or convection (anisotropy in general). As discussed in the main text, at low X-ray luminosities 
the quasi-static shell is capable of removing angular momentum from the rotating
magnetosphere via convective motions. As the accretion rate exceeds some critical value, 
strong Compton cooling causes a free-fall gap to appear above the magnetosphere,
and angular momentum cannot be transferred from the magnetosphere to the shell any more. 

The equation of motion \Eq{v_R1} in the absence of viscosity reads:
\beq{eom}
u_r\frac{du_r}{dR}=-\frac{1}{\rho}\frac{dP_g}{dR}-
\frac{1}{\rho}\frac{dP_\parallel^t}{dR}-\frac{2(P_\parallel^t-P_\perp^t)}{\rho R}-\frac{GM}{R^2}
\eeq
Here $P_g=\rho c_s^2/\gamma$ is the gas pressure, and $P^t$ the pressure due to turbulent pulsations, which in general are anisotropic:
\beq{}
P_\parallel^t =\rho <u_\parallel^2>=\rho m_\parallel^2 c_s^2=\gamma P_g  m_\parallel^2
\eeq
\beq{}
P_\perp^t =2\rho <u_\perp^2>=2\rho m_\perp^2 c_s^2 =2\gamma P_g  m_\perp^2
\eeq
$<u_t^2>=<u_\parallel^2>+2<u_\perp^2>$ is the  
turbulent velocity dispersion, $m_\parallel^2$ and $m_\perp^2$ are the parallel and 
perpendicular turbulent Mach numbers squared). 

From the first law of thermodynamics we have
\beq{1td}
\frac{dE}{dR}=\frac{P_g}{\rho}\frac{d\rho}{dR}+T\frac{dS}{dR}\,,
\eeq
where the specific internal energy (per gram) is 
\beq{E}
E=c_VT=\frac{c_s^2}{\gamma(\gamma-1)}\,,
\eeq
and the heat capacity is  
\beq{cV}
c_V=\frac{{\cal R}}{\mu_m}\frac{1}{\gamma-1}\,.
\eeq
From the second law of thermodynamics, 
the specific entropy change can be written using the rate of the 
specific heat change $dQ/dt$ [erg/s/g] as  
\beq{}
T\frac{dS}{dR}=\frac{dQ}{dR}=\frac{dQ/dt}{u_r}\,.
\eeq
Using the mass continuity equation
\beq{dotM}
\dot M=4\pi R^2\rho u_r\,,
\eeq
we find 
\beq{}
\frac{1}{\rho}\frac{d\rho}{dR}=-\frac{2}{R}-\frac{1}{2u_r^2}\frac{du_r^2}{dR}\,.
\eeq
Using the relation $c_s^2=\gamma{\cal R}T$, we finally obtain:
\beq{dcdR}
\frac{1}{c_s^2}\frac{dc_s^2}{dR}=(\gamma-1)\left[-\frac{2}{R}-\frac{1}{2u_r^2}\frac{du_r^2}{dR}\right]+
\frac{dQ/dt}{u_rc_VT}\,.
\eeq
Note that this equation can also be derived 
directly from the ideal gas equation of state written 
in the form
\beq{eos}
P_g=Ke^{S/c_V}\rho^\gamma\,,
\eeq
where $K$ is some constant.

Using \Eq{dcdR}, the gas pressure gradient
can be rewritten in the form:
\beq{dPdR2}
\frac{1}{P_g}\frac{dP_g}{dR}=\frac{c_s^2}{c_V u_r}\frac{dQ/dt}{T}
+c_s^2\left[-\frac{2}{R}-\frac{1}{2u_r^2}\frac{du_r^2}{dR}\right]
\eeq 
Plugging \Eq{dPdR2} into the equation of motion finally yields:
\beq{eom1}
\frac{1}{2}\frac{1}{u_r^2}\frac{du_r^2}{dR}=
\left[c_s^2(1+\gamma m_\parallel^2)
\left(\frac{2}{R}-\frac{dQ/dt}{c_V u_rT}\right)
-2c_s^2\frac{(m_\parallel^2-m_\perp^2)}{R}-\frac{GM}{R^2}
\right]/\left[u_r^{2^{}}-c_s^2(1+\gamma m_\parallel^2)\right]\,.
\eeq 
Note also that in the strongly anisotropic case where $m_\parallel^2=m_t^2\gg m_\perp^2$, the role of 
turbulence increases in comparison with the isotropic case where $m_\parallel^2=m_\perp^2=(1/3) m_t^2$.

We can also introduce the Mach number in the flow 
${\cal M}\equiv u_r/c_s$. Then from \Eq{dcdR} and
\Eq{eom1} we derive the equation for the Mach number:
\begin{eqnarray}
\label{mach}
&\frac{[{\cal M}^2-(1+\gamma m_\parallel^2)]}{{\cal M}^2}
\frac{d{\cal M}^2}{dR}=\nonumber\\
& \left\{
\frac{2\left[(\gamma-1){\cal M}^2-(\gamma+1)(m_\parallel^2-m_\perp^2)\right]}{R}-
\frac{\left[{\cal M}^2+\gamma(1+\gamma m_\parallel^2)\right]}{c_VT}\frac{dQ}{dR}-
\frac{(\gamma+1)GM}{R^2c_s^2}
\right\}\,,
\end{eqnarray}
where we have substituted $(dQ/dt)=u(dQ/dR)$. Equations (\ref{dcdR}), (\ref{eom1}) and
(\ref{mach}) can be used to solve the dynamics of the accretion flow for pairs of independent variables $(u, c_s)$, $(u,{\cal M})$ or $(c_s, {\cal M})$.  Here, however, we shall
only consider the behaviour of the flux near the singular point. To this end, we can use \Eq{eom1}.

\Eq{eom1} has a singular saddle point where the denominator vanishes:
\beq{saddle}
u_r^2=c_s^2(1+\gamma m_\parallel^2)\,.
\eeq 
So must the numerator, from which we find the quadratic equation for the velocity at the singular point:
\beq{qe}
u_r^2\frac{2}{R}\myfrac{1+(\gamma-1)m_\parallel^2+m_\perp^2}{1+\gamma m_\parallel^2}
-u_r\myfrac{dQ/dt}{c_V T}-\frac{GM}{R^2}=0\,.
\eeq
Remember that in 
the adiabatic case ($dQ/dt=0$) without turbulence 
at the saddle point we have simply 
\beq{cs2ad}
u_r^2=c_s^2=\frac{GM}{2R}\,.
\eeq 
We stress that the presence of turbulence increases the velocity at the singular point. For example, for $\gamma=5/3$ we find for strong anisotropic turbulence $u^2=c_s^2(1+(5/3)m_\parallel^2)$; for the isotropic turbulence the correction is smaller:  $u^2=c_s^2(1+(5/9)m_t^2)$. 
The transition through the sound speed (the sonic point where $u^2=c_s^2$) lies above the saddle point due to turbulence, and there is no singularity in the sonic point. 

First let us determine the turbulence heating rate in the quasi-static shell  $(dQ/dt)^+_t$: 
\beq{}
\myfrac{dQ}{dt}^+_t=\frac{1}{2}\frac{<u_t^2>}{t_t}\,,
\eeq
where the characteristic time of the turbulent heating is 
\beq{}
t_t=\alpha_t\frac{R}{u_t}=\alpha_t\frac{R}{m_t c_s}\,,
\eeq
with $\alpha_t$ being a dimensionless constant characterizing the turbulent dissipation 
energy rate and the turbulent Mach number is $m_t^2\equiv m_\parallel^2+2m_\perp^2$.
The turbulent heating rate can thus be written as
\beq{}
\myfrac{dQ}{dt}^+_t=\frac{c_s^3}{2\alpha_t R}m_t^3\,.
\eeq 

In the case of Compton cooling we have
\beq{}
\myfrac{dQ}{dt}_C^-=-\frac{c_V (T-T_x)}{t_C}\,,
\eeq
where $t_C$ is the Compton cooling time [\Eq{t_comp}]. 

\Eq{qe} can now be written in the form:
\beq{qe1}
u_r^2\frac{2}{R}\myfrac{1+(\gamma-1)m_\parallel^2+m_\perp^2}{1+\gamma m_\parallel^2}
-u_r^2\frac{c_s}{u_r}\frac{\gamma(\gamma-1)m_t^3}{2\alpha_t R}
+\frac{u_r(1-T_x/T)}{\gamma t_C}-\frac{GM}{R^2}=0\,.
\eeq
As we study the accretion problem, the sign of the velocity $u_r=dR/dt$ is negative,
so below we shall write $u_r=-|u_r|$. Then for the absolute value of the velocity at the
singular point where the sound speed is $c_s/|u_r|=-1/(1+\gamma m_\parallel^2)^{1/2}$ 
we have the quadratic equation:
\beq{qe2}
u_r^2\frac{2}{R}\myfrac{1+(\gamma-1)m_\parallel^2+m_\perp^2}{1+\gamma m_\parallel^2}
+u_r^2\frac{1}{(1+\gamma m_\parallel^2)^{1/2}}\frac{\gamma (\gamma-1)m_t^3}{2\alpha_t R}
-\frac{|u_r|(1-T_x/T)}{\gamma t_C}-\frac{GM}{R^2}=0\,.
\eeq 

In this case, the solution  to \Eq{qe} reads:
\beq{u_sol}
|u_r|=\frac{R(1-T_x/T)}{4\gamma t_C A}+\sqrt{\frac{2GM}{R}}
\left[\frac{1}{4A}+\frac{R}{2GM}\frac{R^2(1-T_x/T)^2}{16\gamma^2t_C^2A^2}\right]^{1/2}\,,
\eeq
where we have introduced the dimensionless factor 
\beq{A}
A=\frac{1+(\gamma-1)m_\parallel^2+m_\perp^2}{1+\gamma m_\parallel^2}+
\frac{\gamma (\gamma-1)( m_\parallel^2+2m_\perp^2)^{3/2}}{4\alpha_t (1+\gamma m_\parallel^2)^{1/2}}\,.
\eeq
In the case of isotropic turbulence where  
$m_\parallel=m_\perp=1/\sqrt{3}, m_t=1$, for $\gamma=5/3$ the factor $A\approx 1.23$, and in the case of strongly anisotropic turbulence where
$m_\parallel=1, m_\perp=0, m_t=1$, this factor is $A\approx 0.8$.
 
In units of the free-fall velocity the solution \Eq{u_sol} reads:
\beq{cs1}
f(u)=\frac{|u_r|}{u_{ff}}=\frac{(1-T_x/T)}{4\gamma A}\myfrac{t_{ff}}{t_C}+
\frac{1}{2}
\left[\frac{1}{A}+\frac{(1-T_x/T)^2}{4\gamma^2A^2}\myfrac{t_{ff}}{t_C}^2\right]^{1/2}\,.
\eeq 

With Compton cooling present, the temperature changes exponentially:
\beq{}
T=T_x+(T_{cr}-T_x)e^{-t/t_C}
\eeq
see the main text).  
When cooling is slow, $t_{ff}/t_C\ll 1$, 
the critical point lies inside the Alfv\'en surface, i.e. no transition through the
critical point occurs in the flow before it meets the magnetosphere, and in this case 
we expect settling accretion. If this point lies above the Alfv\'en surface, the velocity of the flow
may become supersonic above the magnetosphere, and one may thus expect the
formation of a shock. Both turbulence and rapid cooling shifts
the location of the critical point upwards in the flow. 

In the case of rapid cooling $t_{ff}/t_C\gg 1$, 
$T\to T_x$, so again $u_r/u_{ff}\approx 1/2$ (cf. \Eq{cs2ad} for an adiabatic flow), but the critical point now lies above the Alfv\'en surface, so a free-fall gap above the magnetosphere 
appears. The ratio $f(u)=|u_r|/u_{ff}$ reaches a maximum at $t_{ff}/t_C\approx 0.46$ (assuming a typical ratio $T_{cr}/T_x=10$), and depending on the value of $A=0.8\div 1.23$ (anisotropic or isotropic turbulence) 
it equals to $f(u)=0.5-0.6$.  
  
\section*{Acknowledgements}
The authors would like to thank Dr. V. Doroshenko (IAAT) for courtesly providing
the torque-luminosity plots for GX 301-2 and Vela X-1, and Dr. V. Beskin and Dr. V. Suleimanov for discussions. NIS thanks the Max-Planck Institute for Astrophysics (Garching) for shown hospitality.
The work by NIS, KAP and AYK is supported by RFBR grants 
09-02-00032, 12-02-00186 and 10-02-00599.  LH is supported
by a grant from the Wenner-Gren foundations (Sweden).


\begin{thebibliography}{99}
 
\bibitem
{Bildsten_ea97} 
Bildsten L et al \protect\emph{ Astrophys. J. Suppl.} \textbf{113} 367(1997) 
 
 
 
\bibitem
{ShakuraSunyaev73} Shakura N I, Sunyaev R A \textit{Astron. Astrophys.} \textbf{24} 337 (1973) 
 
 
\bibitem
{PringleRees72} Pringle J E, Rees M J \textit{Astron. Astrophys.} \textbf{21} 1 (1972) 
 
\bibitem
{GhoshLamb79}
Ghosh P, Lamb F K \textit{Astrophys. J.} \textbf{234} 296 (1979) 
 
 
\bibitem
{Lovelace_ea95} Lovelace R V E, Romanova M M, Bisnovatyi-Kogan G S \textit{Mon. Not. R. Astron. Soc.} \textbf{275} 244 (1995) 
 
 
\bibitem
{KluzniakRappaport07}
Klu\'zniak W, Rappaport S \textit{Astrophys. J.} \textbf{671} 1990 (2007)
 
\bibitem
{FryxellTaam88} Fryxell B A, Taam R E \textit{Astrophys. J.} \textbf{335} 862 (1988) 
 
\bibitem
{Ruffert97}
Ruffert M \textit{Astron. Astrophys.} \textbf{317} 793 (1997) 
 
\bibitem
{Ruffert99}
Ruffert M \textit{Astron. Astrophys.} \textbf{346} 861 (1999)
 
\bibitem
{Burnard_ea83} Burnard D J, Arons J, Lea S M \textit{Astrophys. J.} \textbf{266} 175 (1983)
 
\bibitem
{DaviesPringle81} Davies R E, Pringle J E \textit{Mon. Not. R. Astron. Soc.} \textbf{196} 209 (1981) 
 
\bibitem
{IllarionovKompaneets90} Illarionov A F, Kompaneets D A \textit{Mon. Not. R. Astron. Soc.} \textbf{247} 219 (1990) 
 
 
\bibitem
{BisnovatyiKogan91} Bisnovatyi-Kogan G S \textit{Astron. Astrophys.} \textbf{245} 528 (1991)
 
\bibitem
{Shakura_ea12} Shakura N I, Postnov K A, Kochetkova A Yu, Hjalmarsdotter L \textit{Mon. Not. R. Astron. Soc.} \textbf{420} 216  (2012)
 
\bibitem
{IllarionovSunyaev75} Illarionov A F, Sunyaev R A \textit{Astron. Astrophys.} \textbf{39} 185 (1975)
 
 
\bibitem
{ElsnerLamb77} Elsner R F, Lamb F K \textit{Astrophys. J.} \textbf{215} 897 (1977) 
 
 
\bibitem
{AronsLea76a} Arons J, Lea S M  \textit{Astrophys. J.} \textbf{207} 914 (1976)
 
 
\bibitem
{Bondi52}
Bondi H \textit(Mon. Not. R. Astron. Soc.) \textbf{112} 195 (1952) 
 
 
\bibitem
{Kompaneets56} Kompaneets A S \textit{ZhETP} \textbf{31} 876 (1956) 
 
\bibitem
{Weymann65} 
Weymann R \textit{Phys. Fluids} \textbf {8} 2112 (1965) 
 
 
\bibitem
{Shakura_ea12b}
Shakura N I, Postnov K A, Hjalmarsdotter L \textit{Mon. Not. R. Astron. Soc.} \textbf{428} 670 (2013)
 
\bibitem
{Doroshenko_ea11}
Doroshenko V,  Santangelo A, Suleimanov V \textit{Astron. Astrophys.} \textbf{529} 52 (2011)
 
 
\bibitem
{Finger_ea11} Finger M et al \textit{\hbox{http://gammaray.nsstc.nasa.gov/gbm/science/pulsars/lightcurves/gx1p4.html}}
 
 
\bibitem
{Ikhsanov12} 
Ikhsanov N R, Beskrovnaya N G \textit{Astron. Rep.} \textbf{56} 589 (2012)
 
\bibitem
{Lipunov87}
Lipunov V M \textit{Astrophysics of Neutron Stars} (Berlin: Springer, 1992)
 
\bibitem
{Chakrabarty_ea97} 
Chakrabarty D et al \textit{Astrophys. J. Lett.} \textbf{481} L101 (1997) 
 
\bibitem
{Sunyaev78} Sunyaev R A   in: \textit{Physics and astrophysics of neutron stars and black holes} (Amsterdam: North Holland Publ., 1978) p. 697 
 
 
\bibitem
{Ho_ea89} Ho C et al \textit{Mon. Not. R. Astron. Soc.} \textbf{238} 1447 (1989) 
 
\bibitem
{Nelson_ea97}
Nelson R W et al \textit{Astrophys. J.} \textbf{488} L117 (1997) 
 
 
\bibitem
{AronsLea76b} Arons J, Lea S M \textit{Astrophys. J.} \textbf{210} 792 (1976)
 
 
\bibitem
{Hunt71} 
Hunt R \textit{Mon. Not. R. Astron. Soc.} \textbf{154} 141 (1971) 
 
 
\bibitem
{GonzalezGalan_ea12} Gonz\'alez-Gal\'an A et al \textit{Astron. Astrophys.} \textbf{537} A66 (2012)
 
\bibitem
{ChashkinaPopov12} Chashkina A A, Popov S B \textit{New Astron.} \textbf{17} 594 (2012)
 
\bibitem
{Lue_ea12} 
Lue G-L et al \textit{Mon. Not. R. Astron. Soc.} \textbf{424} 2265 (2012)
 
 
 
\bibitem
{Lutovinov_ea12} Lutovinov A, Tsygankov S, Chernyakova M
\textit{Mon. Not. R. Astron. Soc.} \textbf{423} 1978 (2012)
 
 
\bibitem
{Marcu_ea11} Marcu D M et al \textit{Astrophys. J.} \textbf{742} L11 (2011)
 
\bibitem
{PopovTurolla12} Popov S B, Turolla R \textit{Mon. Not. R. Astron. Soc.} \textbf{421} L127 (2012)
 
 
\bibitem
{Fu12} Fu L, Li X-D, \textit{Astrophys. J.} \textbf{757} id171 (2012)
 
 
 
\bibitem
{Koh_ea97} Koh D T et al \textit{Astrophys. J.} \textbf{479} 933 (1997) 
 
\bibitem
{White_ea76} White N E et al
\textit{Astrophys. J.} \textbf{209} L119 (1976) 
 
 
\bibitem
{Kaper_ea06} Kaper L, van der Meer A, Najarro F \textit{Astron. Astrophys.} \textbf{457} 595 (2006) 
 
 
\bibitem
{deKoolAnzer93} de Kool M, Anzer U \textit{Mon. Not. R. Astron. Soc.} \textbf{262} 726 (1993) 
 
\bibitem
{Nagase89} Nagase F \textit{Publ. Astron. Soc. Jap.} \textbf{41} 1 (1989) 
 
\bibitem
{PravdoGhosh01} Pravdo S H, Ghosh P \textit{Astrophys. J.} \textbf{554} 383 (2001)
 
 
\bibitem
{LaBarbera_ea05} La Barbera A et al
\textit{Astron. Astrophys.} \textbf{438} 617 (2005) 
 
 
\bibitem
{Kreykenbohm_ea04} Kreykenbohm I et al
\textit{Astron. Astrophys.} \textbf{427} 975 (2004) 
 
 
\bibitem
{Doroshenko_ea10} Doroshenko V et al \textit{Astron. Astrophys.} \textbf{515} A10 (2010) 
 
\bibitem
{Quaintrell_ea03} 
Quaintrell H et al \textit{Astron. Astrophys.} \textbf{401} 313 (2003) 
 
\bibitem
{vanKerkwijk_ea95} 
van Kerkwijk M H et al \textit{Astron. Astrophys.} \textbf{303} 483 (1995) 
 
 
\bibitem
{Rappaport75} Rappaport S \textit{IAU Circ.} \textbf{2869} 2 (1975) 
 
\bibitem
{Bochkarev_ea75} 
Bochkarev N G, Karitskaja E A, Shakura N I \textit{Sov. Astron. Lett.} \textbf{1} 237 (1975)
 
\bibitem
{Nagase_ea86} Nagase F et al
\textit{Publ. Astron. Soc. Jap.} \textbf{38} 547 (1986) 
 
\bibitem
{Watanabe_ea06}
Watanabe S et al \textit{Astrophys. J.} \textbf{651} 421 (2006) 
 
 
\bibitem
{Staubert_03} Staubert R \textit{Chin.
 J. Astron. Astrophys. Suppl.} \textbf{3} 270 (2003) 
 
 
\bibitem
{Doroshenko11} Doroshenko V \textit{PhD Thesis University of Tuebingen (IAAT)} (2011)
 
\bibitem
{Davidsen_ea77} Davidsen A, Malina R, Bowyer S \textit{Astrophys. J.} \textbf{211} 866 (1977) 
 
 
\bibitem
{Hinkle_ea06} Hinkle K H, Fekel F C, Joyce R R, Wood P R, Smith V V, Lebzelter T \textit{Astrophys. J.} \textbf{641} 479 (2006) 
 
\bibitem
{Makishima_ea88} Makishima K et al \textit{Nature} \textbf{333} 746 (1988) 
 
 
\bibitem
{Dotani_ea89} Dotani T et al
\textit{Publ. Astron. Soc. Jap.} \textbf{41} 472 (1989) 
 
 
 
\bibitem
{HenaultBrunet_ea12} H\'enault-Brunet V et al \textit{Mon. Not. R. Astron. Soc.}
\textbf{420} L13 (2012)
 
\bibitem
{Haberl_ea12} Haberl F et al \textit{Astron. Astrophys.} \textbf{537} L1 (2012)
 
\bibitem
{PopovTurolla} Popov S B, Turolla R \textit{Mon. Not. R. Astron. Soc.}
\textbf{421} L127 (2012)
 
\bibitem
{FuLi12} Fu Lei, Li Xiang-Dong \textit{Astrophys. J.}
\textbf{757} 171 (2012)
 
 
\bibitem
{Reig_ea12} Reig P, Torrej\'on J M, Blay P \textit{Mon. Not. R. Astron. Soc.} \textbf{425} 529 (2012)
 
 
\bibitem{Ribo_ea06}
Rib\'o M et al.  \textit{Astron. Astrophys.} \textbf{449} 687 (2006)
 
\bibitem
{Torrejon_ea04}
Torrej\'on et al. \textit{Astron. Astrophys.} \textbf{423} 301 (2004)
 
\bibitem
{Masetti_ea04}
Masetti N et al. \textit{Astron. Astrophys.} \textbf{423} 311 (2004) 
 
\bibitem
{Blay_ea06}
Blay P et al. \textit{Astron. Astrophys.} \textbf{438} 963 (2005)
 
\bibitem
{Wang09}
Wang W \textit{Mon. Not. R. Astron. Soc.} \textbf{398} 1428 (2009)
 
 
\bibitem
{Raymond_ea76} Raymond J C, Cox D P, Smith B W \textit{Astrophys. J.} \textbf{204} 290 (1976) 
 
\bibitem
{Cowie_ea81} Cowie L L, McKee C F, Ostriker J P \textit{Astrophys. J.} \textbf{247} 908 (1981)  
 
\bibitem
{Tarter_ea69} Tarter C B, Tucker W H, Salpeter E E \textit{Astrophys. J.} \textbf{156} 943 (1969) 
 
 
\bibitem
{Hatchett_ea76} 
Hatchett S, Buff J, McCray R \textit{Astrophys. J.} \textbf{206} 847 (1976) 
 
\bibitem
{SunyaevShakura77} Sunyaev R A, Shakura N I \textit{Sov. Astron. Lett.} \textbf{3} 138 (1977) 
 
 
\bibitem
{Marykutty_ea10} 
Marykutty J et al
\textit{Mon. Not. R. Astron. Soc.} \textbf{407} 285 (2010) 
 
 
\bibitem
{Ducci_ea10} 
Ducci L, Sidoli L, Paizis A \textit{Mon. Not. R. Astron. Soc.} \textbf{408} 1540 (2010) 
 
 
\bibitem
{LandauLifshitz86} Landau L D, Lifshitz E M \textit{Fluid Mechanics}(Pergamon Press, 1959)
 
 
\bibitem
{Wasiutinski46}
Wasiuty\'nski J \textit{Studies in Hydrodynamics and Structure of Stars and Planets} (Oslo, 1946) 
 
 
\bibitem
{ShakuraSunyaev88} 
Shakura N I, Sunyaev R A \textit{Adv. Space Res.} \textbf{8} 135 (1988) 
 
\bibitem
{Shakura_ea77}
Shakura N I, Sunyaev R A, Zilitinkevich S S \textit{Astron. Astrophys.} \textbf{62} 179 (1978)
 
 
 
\bibitem
{Parker63} Parker E 
\textit{Interplanetary dynamical processes} (New York: Interscience Publ., 1963) 
 
\bibitem
{Beskin05} Beskin V S
\textit{MHD Flows in Compact Astrophysical Objects: Accretion, Winds and Jets} (Berlin: Springer, 2010) 
 
 
\end{thebibliography}
\end{document}